\newif\ifMNRAS
\MNRASfalse 

\PassOptionsToPackage{pdfpagelabels=false}{hyperref} 

\ifMNRAS
    \documentclass[fleqn,usenatbib,useAMS]{mnras}
    \usepackage{graphicx}
\else

\documentclass[]{aastex63}
\fi

\usepackage{hyperref}
\usepackage{amsmath,amsfonts,amssymb}
\usepackage{mathrsfs}
\usepackage{mathtools}
\usepackage[utf8]{inputenc}

\usepackage[normalem]{ulem}

\usepackage{pifont}

\newcommand{\adam}[1]{{#1}}

\date{Accepted May 23, 2025}

\shorttitle{BH CCSNe}

\begin{document}
\title{Channels of Stellar-mass Black Hole Formation} 
\correspondingauthor{Adam Burrows}
\email{burrows@astro.princeton.edu}
\author[0000-0002-3099-5024]{Adam Burrows}
\affiliation{Department of Astrophysical Sciences, Princeton University, NJ 08544, USA; School of Natural Sciences, Institute for Advanced Study, Princeton, NJ 08540}
\author[0000-0002-0042-9873]{Tianshu Wang}
\affiliation{Department of Astrophysical Sciences, Princeton University, NJ 08544, USA}
\author[0000-0003-1938-9282]{David Vartanyan}
\affiliation{Carnegie Observatories, 813 Santa Barbara St., Pasadena, CA 91101, USA; NASA Hubble Fellow}

\begin{abstract}
On the basis of a large collection of detailed 3D core-collapse supernova simulations carried to late times, we identify four channels of stellar mass black hole formation. Our examples for Channel 1 can involve the formation of black holes in energetic asymmetric supernova explosions. Our Channel 2 example involves a modest supernova explosion that nevertheless leaves behind a black hole. The latter may not be easily distinguishable from ``standard" supernovae that birth neutron stars. Our Channel 3 example experiences an aborted core-collapse explosion, more often in the context of a low-metallicity progenitor, whose residue is a black hole with a mass perhaps up to $\sim$40 $M_{\odot}$. The latter may be accompanied by a pulsational-pair instability supernova (PPISN). Channel 4 is the only quiescent or ``silent" scenario for which perhaps $\sim$5 to 15 $M_{\odot}$ black holes are left. Where appropriate, we estimate $^{56}$Ni yields, explosion energies, approximate recoil speeds, and residual black hole masses. The progenitor mass density and binding energy profiles at collapse influence the outcome in a systematic way. \adam{We speculate that} the statistics and prevalence of these various channels depend not only on still evolving supernova theory, but on remaining issues with the theory of massive star evolution, binary interaction, wind mass loss, metallicity, and the nuclear equation of state. Importantly, we suggest, but have not proven, that the silent channel for black hole formation may not be the dominant formation modality. 
\end{abstract} 

\ifMNRAS
\else 
\fi

\section{Introduction}
\label{sec:int}  
   
Stellar-mass black holes in high-mass X-ray binaries (HMXBs) and low-mass X-ray binaries (LMXBs) have been studied in the galaxy for decades \citep{mcclintock2006,2023arXiv230409368M}.  They have measured masses (with modest error bars) between $\sim$4 and $\sim$21 $M_{\odot}$, with the peak near $\sim$8 $M_{\odot}$ \citep{ozel2010,farr2011}. In addition, the Gaia astrometric mission has recently found three non-accreting black holes with masses of $\sim$9 $M_{\odot}$ \citep[BH1 and BH2,][]{elbadry_1_2023,elbadry_giant_2023} and $\sim$33 $M_{\odot}$ \citep[BH3][]{gaia3_2024}, with many more promised. \adam{These black hole masses are likely to be very close to their birth masses, since subsequent accretion in the associated wider binary context is likely to be minimal \citep{cappelluti2024}. However, the mass added after black hole birth when in a tight binary can vary from a negligible amount to perhaps $\sim$0.5 $M_{\odot}$ \citep{Xing2025}, but could be larger \citep{klencki2025}. In this paper, we deal exclusively with black hole birth masses.} Suggestively, the lower-mass BH1 and BH2 orbit approximately solar-metallicity stars, while the more massive Gaia black hole BH3 orbits a low-metallicity ($[Fe/H]$ $\sim$ $-$2.5) star.
Also intriguing is the fact that all three of these black holes    reside in highly eccentric ($\sim$0.4, $\sim$0.5, and $\sim$0.7, respectively) orbits.   


However, the recent epochal detections by the LIGO/Virgo/KAGRA (LVK) collaboration \citep{aligo,ligo_3_2023,virgo,kagra} of gravitational waves from scores of merging black hole binaries have catapulted the study of the origin, prevalence, and mass function of such black holes to the forefront of astrophysical research \citep[][and references therein]{li2024}. All these black holes are born through the collapse of the cores of massive stars, or perhaps the merger of black holes formed through such a channel, but which progenitor star for a given metallicity gives birth to stellar-mass black holes is very much a topic of ongoing research. Interestingly, the LVK collaboration has identified in the putative lower mass gap \citep{shao2022} a possible black hole in the mass interval $\sim$2.5$-$4.5 $M_{\odot}$ \citep[GW230529,][]{Abac_2024}. The basic questions remain: What are the detailed stellar contexts of black hole birth and the true galactic black hole mass distribution function? The number of stellar-mass black holes in the galaxy may be \adam{as many as 10$^8$ \citep{Kaczmarek2025},} and this may complement an inferred 10$^8$ galactic neutron stars, but we don't yet know.  

The terminal evolutionary phase of a massive star with ZAMS\footnote{ZAMS: ``Zero Age Main Sequence"} mass greater than or equal to approximately 8 M$_{\odot}$ leads to the creation of either a neutron star or a stellar mass black hole. The creation of a neutron star in such a context must be accompanied by a core-collapse supernova (CCSN) explosion. However, whether the same is true of stellar mass black hole birth now seems to be a more nuanced issue. In the past, the astronomy community speculated that all stellar-mass black hole formation would be ``silent," with no supernova display \citep[an ``unnova,"][]{2008ApJ...684.1336K,gerke2015,adams2017a,adams2017b,beasor2024,de2024}.  Furthermore, the implication was that the stalled shock was not in these cases even temporarily revived. However, the emerging modern theory of core collapse is now suggesting something different.  

For a 40-M$_{\odot}$ zero-metallicity progenitor \citep{2010IAUS..265....3W}, \citet{chan2018} and \citet{moriya} witnessed a weak explosion that was accompanied by black hole birth due to significant late-time fallback of initially ejected matter. Using approximate neutrino transport and exploring initially rotating and non-rotating variants of a solar-metallicity 40-M$_{\odot}$ progenitor \citep{sukhbold2018}, \citet{2021ApJ...914..140P} concluded that their rotating models exploded and that the slowly-rotating and non-rotating models left black holes. The rapidly-rotating model left a neutron star. Using the same solar-metallicity 40-$M_{\odot}$ progenitor, \citet{ott2018_rel} found that though the core mantle initially exploded, it did so weakly and that a black hole would ultimately form without much of an external display. However, they and these other researchers calculated no model beyond $\sim$1 second after bounce, a time that has been determined to be too short to   determine the asymptotic state of core collapse \citep{muller2016,burrows_2020,bollig2021,burrows_correlations_2024}.
Moreover, \cite{2020PhRvD.101l3013W}, \cite{2018MNRAS.477L..80K}, and \citet{2023arXiv230706192K} observed that massive stars with ZAMS masses of 40-$M_{\odot}$ to 75-$M_{\odot}$ could launch the stalled shock, but that black hole birth rapidly followed. The short interval between explosion and black hole birth ultimately aborted neutrino heating and undermined the supernova into infall, further fattening the black hole {\footnote{\adam{The reader is referred to \citet{burrows_40}, section 2 there, for a more in-depth discussion and critique of 3D black hole formation models of other groups, with an emphasis on the 40-$M_{\odot}$ solar-metallicity progenitor.}} However, \citet{sykes_muller_BH} in a series of recent 2D simulations of zero-metallicity stars from 60 to 90 $M_{\odot}$ find that they explode with canonical supernova energies, while leaving behind black holes with masses from $\sim$21 to $\sim$34 $M_{\odot}$. These could be examples of so-called ``fallback" supernovae and black hole formation (see \S\ref{bh_gap_weak}).
Finally, we \citep{burrows_40} recently simulated in 3D the solar-metallicity 40-$M_{\odot}$ progenitor model of \citet{sukhbold2018} using the SFHo \citep{2013ApJ...765L...5S} nuclear equation of state (EOS) and found using our state-of-the-art code F{\sc{ornax}} that it exploded vigorously and quite aspherically. 

Hence, what has emerged from these disparate recent publications is a picture of stellar-mass black hole formation that deviates from the simple picture often presented. There is even the recent inference of a supernova remnant around the black hole micro-quasar SS 433 \citep{ss433}. Black hole formation could be accompanied by a strong supernova explosion, a weak supernova explosion, or an aborted explosion, and not just (perhaps) the quiescent dud of common conjecture. The progenitor mass, metallicity \citep{belczynski2002,belczynski2010}, stellar mass loss prescription \citep{smartt2009b}, and binary interactions are all factors in the outcome, since these have a bearing on the total stellar mass that remains at the final evolutionary stage, the binding energy of the stellar envelope to be overcome, and the ``compactness" \citep{2011ApJ...730...70O} of the inner core where supernova dynamics is determined\footnote{The compactness parameter loosely characterizes the inner core structure and is defined as
\begin{equation}     
\xi_M= \frac{M/M_{\odot}}{R(M)/1000\, \mathrm{km}}\,,
\end{equation}
where the subscript $M$ denotes the interior mass coordinate at which this parameter is evaluated. Generally when exploring the viability of a standard CCSN explosion we have evaluated the compactness parameter at $M$ = 1.75 M$_{\odot}$.  Originally, compactness was introduced to study black hole formation and was calculated at 2.5 $M_{\odot}$.}. Generally, low metallicity makes a vigorous explosion at very high compactness difficult and encourages the formation of a more massive black hole. In addition, the nuclear equation of state determines the maximum mass of a proto-neutron star (PNS). As such, it too has an influence on how long neutrino heating can continue after shock revival before black hole formation truncates it and, therefore, on the ultimate explosion energy, the ejected mass, and the final black hole mass.  

Importantly, despite recent conceptual and numerical progress in core collapse theory, the expected distribution function of black hole masses that the population of massive stars produces and the mapping of initial progenitor to final residual black hole mass are still very much in flux. CCSN theory, though it has entered a sophisticated phase, is not yet as predictive as observers need. The terminal phase of massive star evolution includes burning shell mergers and mixing that are only now being studied hydrodynamically \citep{Chatzopoulos2016,muller2016,muller2017,Muller2019,takahashi2019,fields2020,Fields2021,2021MNRAS.506L..20Y,2021ApJ...908...44Y}. The different 1D massive star stellar evolution models \citep{swbj16,sukhbold2018,limongi2024,laplace2024} have not converged, though there has of late been considerable progress on that front. Moreover, binary interaction must assume an important role \citep{woosley2019,laplace2024}, particularly in the resulting black hole mass function, but perhaps also in the unstable core structures themselves \citep{laplace2021}. 

Despite this, in the absence of definitive guidance from evolving CCSN theory and the lack of a converged understanding of massive star evolution, several simple prescriptions have leapt into the breach \citep{zhang2008,2011ApJ...730...70O,2012ApJ...757...69U,muller2016,2016ApJ...818..124E,2016MNRAS.460..742M,swbj16,2019ApJ...870....1E,mandel2021,schneider2023,temaj2024} purporting to provide a recipe predicting whether a given progenitor core structure (often crudely associated with compactness) will result in a black hole. However, these maps are likely quite imperfect \citep{muller2016,2020ApJ...890..127C,burrows_2020,2021Natur.589...29B,wang,tsang2022,burrows_40,boccioli2024}.  Importantly, there is little nuance in these prescriptions and there are important 3D effects that are generally ignored. Moreover, the majority of these approaches merely assuming a ``silent" outcome. It is the latter with which we take exception in this paper. We find in our detailed 3D simulation investigations to late times after bounce several modalities and channels of black hole formation, some explosive, and this heterogeneity, if true, has a bearing on all aspects of black hole formation. In this paper we explore exemplars of the different routes to black hole formation we have witnessed through our recent 3D numerical simulation campaigns. We will not attempt, however, to generate theoretical distribution functions of final black hole masses, explosion energies and nucleosynthesis (if relevant), kick speeds, or induced spins, though for each channel we do suggest parameter ranges. We feel in the context of CCSN theory that such an exercise, however desirable, is premature. However, we suggest that the story to emerge is a richer and more compelling narrative of black hole birth that needs to be recognized more broadly if true progress is to be made on the birth of stellar-mass black holes. 

In \S\ref{method}, we summarize the general computational methods we have employed.  Then, in \S\ref{basics}, we provide an aside on some of the salient and relevant aspects of current core-collapse supernova theory that have a bearing on the questions at hand and aren't generally appreciated.  Then, in \S\ref{channels} we list and describe the four different modalities we observe in our 3D CCSN simulations that lead to a stellar-mass black hole and some of the general physical and dynamical characteristics of each channel.  We identify whether black hole formation is accompanied by an explosion, the mass of the residue, the net recoil kicks expected, and the explosive nucleosynthesis (if any). We also distinguish the gravitational-wave and neutrino signatures of each path. The shallowness of the density profile exterior to the inner core, loosely indexed by the ``compactness," will be seen to play a major role in each scenario.  Finally, in \S\ref{conclusions} we summarize the resulting more nuanced and variegated landscape for stellar-mass black hole formation that is collectively suggested.
 
\section{Method}
\label{method}

We employ our workhorse code F{\sc{ornax}} to simulate the dynamical collapse and explosion phase to many seconds after bounce, starting with initial progenitor models created by {\it KEPLER} \citep[][and S. Woosley, private communication]{woosley2002,swbj16,sukhbold2018}.  F{\sc{ornax}} has been exercised for the last eight years on many High-Performance-Computing (HPC) platforms and numerous publications on a wide spectrum of topics in CCSN theory have resulted \citep{radice2017b,burrows2018,vartanyan2018a,vartanyan2018b,vsg2018,burrows_2019,radice2019,hiroki_2019,vartanyan2019,Nagakura2020,vartanyan2020,burrows_2020,nagakura2021,2021Natur.589...29B,2022MNRAS.510.4689V,coleman,burrows_40,burrows_correlations_2024}. A discussion of the numerical methods and microphysics employed can be found in \citet{skinner2019}, \citet{burrows_40}, and in the appendix to \citet{vartanyan2019}. For all 3D F{\sc{ornax}} simulations, we use the SFHo nuclear equation of state \citep{2013ApJ...774...17S}, our default spatial gridding ($1024\times128\times256$: $r\times\theta\times\phi$), and twelve neutrino energy groups for each of three species ($\nu_e$, $\bar{\nu}_e$, and ``$\nu_{\mu}$" [$\equiv$ $\nu_{\mu}$, $\bar{\nu}_{\mu}$, $\nu_{\tau}$, $\bar{\nu}_{\tau}$]). \adam{After black hole formation (when the lapse iteration does not converge, and the code crashes due to the inability to find a stable solution consistent with general relativity (see below),} to continue a simulation we map into either the hydrodynamic component of F{\sc{ornax}} with a point mass and diode inner boundary condition or the FLASH code \citep{FLASH2000} with a similar inner boundary and point mass \citep{vartanyan_breakout_2024}. For all post-F{\sc{ornax}} simulations, we used the Helmholtz equation of state \citep{timmes1999}. For the late-time simulation of the 23-$M_{\odot}$ model, which did not immediately form a black hole, we continued the F{\sc{ornax}} run from 6.22 seconds after bounce with a point mass and a wind inner boundary condition \citep{wong2015} at 500 kilometers (km), and then at 50 seconds after bounce mapped into FLASH and continued it for many physical days.
\adam{The wind was constructed from the flow and its properties actually witnessed in the corresponding F{\sc{ornax}} simulation and its time behavior was extrapolated using a power law fit \citep[see,][]{wang_ti44}. The juncture was exterior to the sonic point of the flow. In this way there is a smooth hydrodynamic transition at the code hand-off, with no artificial shocks or discontinuities.}

\adam{F{\sc{ornax}} incorporates approximate general relativity, with gravitational redshifts applied to the radiation fields \citep{vartanyan2019}.  The monopole term of the gravitational field is the same as in the relativistic TOV equation, with the neutrino energy density and pressure, as well as the kinetic term, included as described in \citet{marek2006}, Case A. In this way, if the core were perfectly spherical the effect of general relativity on the quasi-hydrostatic inner structure would be accurate. The lapse function is derived from this function and directly yields the redshifts. An iterative calculation is required to obtain the gravitational masses and this iteration does not converge when the core is unstable to gravitational collapse. This is also the approach used in \citet{2020PhRvD.101l3013W}. Plots of the lapse at an early and terminal phase can be found in \citet{burrows_40}.} 

The developed 3D FLASH capability \citep{vartanyan_breakout_2024} allows simulating, if necessary, to hours and days after explosion to capture late-time fallback accretion and the evolution of the residual black hole mass. For the FLASH runs, we apply a periodic boundary condition in $\phi$ and use a spatial resolution of $2048\times{192}\times{384}$ to ensure sub-degree angular resolution and a nearly constant radial resolution $\Delta{r}/r$ of 4.7$\times$10$^{-3}$ to 6.7$\times$10$^{-3}$. To ease the Courant timestep condition along the poles, in the FLASH runs we excised five degrees in the northern and southern directions, with a reflective boundary condition in $\theta$. For the post-black-hole formation simulation of the 100-$M_{\odot}$ model, we employed FLASH and used the diode boundary condition at 100 kilometers and an outflow boundary condition at 9.604$\times$10$^{7}$ kilometers. The outer boundary was chosen at a radius where the resolution of the {\it KEPLER} model coarsens due to the ejection of the three pulses associated with the PPISN. Some of the results reported in this paper have been published elsewhere \citep{burrows_2020,vartanyan2023,burrows_40,burrows_kick_2024,burrows_correlations_2024}, but the 19.56-$M_{\odot}$,  100-$M_{\odot}$, and late-time 23-$M_{\odot}$ simulations are new and for all models several new quantities and signatures are provided here for the first time.

\section{Relevant Theory of Neutrino-Driven Core-Collapse Supernova Explosions}
\label{basics}
   
At the end of the quasi-static evolution of stars more massive than perhaps $\sim$8 $M_{\odot}$ (which may last a few to $\sim$30 million years), their degenerate cores eventually achieve the effective Chandrasekhar mass of from $\sim$1.3 to $\sim$2.1 $M_{\odot}$ and gravitationally collapse\footnote{The ``Chandrasekhar" mass that becomes unstable and collapses is a function of the entropy and electron-fraction profiles achieved in the core. More massive stars evolve more quickly, thereby retaining higher core entropies that have less time to decrease due to thermal neutrino emission.  This translates into more massive cores, supported by higher entropies, for more massive progenitor stars.}.  Within hundreds of milliseconds of instability and the onset of implosion, the central mass densities achieve values in excess of that of the atomic nucleus, the matter stiffens to near incompressibility, and the inner core bounces violently into the outer, still infalling, core mantle. At the interface, near a sonic point, a spherical shock wave is formed and launched outward. However, due predominantly to electron capture and the associated copious electron neutrino ($\nu{_e}$) energy emission behind the compressing shock wave in a ``breakout neutrino burst"  (and secondarily to shock photodissociation of the infalling nuclei), the shock stalls within tens of milliseconds into an accretion shock.  Much theoretical work over the last decades has gone into understanding how this stalled shock is revived into explosion. The key for most models is the attendant neutrino-driven convection and turbulence behind the shock, whose primary role is to contribute turbulent stress to augment the pressure behind the shock \citep{bhf1995,janka2012,burrows_2020}.  Without this 3D ingredient most models do not explode in spherical (1D) symmetry (for exceptions, see \citet{kitaura2006} and \citet{wang_low_2024}). At a critical time, determined in a complicated way by the competition between neutrino heating and the augmented stress behind the shock and the ram pressure of the infalling matter \citep{wang}, the structure becomes unstable (reaches a ``critical point") \citep{burrowsgoshy1993} and the shock is (re)launched.  This can take between $\sim$50 and $\sim$1000 milliseconds after core bounce. The accretion of an abrupt silicon/oxygen interface can also kick a model into explosion, since upon encountering the interface the shock experiences an abrupt drop in ram pressure, while maintaining for the accretion time to the core the powering neutrino luminosity.

The scenario articulated above is more or less the standard narrative and is, in broad outline, what all researchers performing detailed 3D CCSN simulations \citep{lentz:15,burrows_2019,vartanyan2019,muller_lowmass,stockinger2020,burrows_2020,bollig2021,sandoval2021,nakamura2022,vartanyan2023,burrows_correlations_2024} see and agree upon. However, there are other aspects of the explosion phenomenon that are more subtle and that are not widely appreciated, yet of central importance to the questions at hand.  

First, we note that many have incorrectly invoked ``compactness" as an ``explodability" condition \citep[e.g.,][]{swbj16}, with the suggestion that a low compactness (steep progenitor core mass density profile) is necessary for explosion by the neutrino mechanism. It isn't, and models at both low and high compactness can explode. In fact, all our models with high compactness experience an initial (re)ignition of the stalled shock. Moreover, recent detailed simulations \citep{burrows_2019,vartanyan2019,muller_lowmass,stockinger2020,burrows_2020,bollig2021,sandoval2021,nakamura2022,vartanyan2023} and general theory \citep{1985ApJ...295...14B,burrows:95,janka2012,burrows2013,burrows_2020} demonstrate that in the context of the neutrino heating mechanism of CCSNe a high compactness is necessary to achieve the canonical supernova energies of $\sim$1.0 Bethe \citep{burrows_correlations_2024,bollig2021}. High compactnesses are associated with high mass accretion rates ($\dot{M}$), that result in both high neutrino accretion luminosities and high neutrino absorption ``optical depths" behind the stalled shock; the product ($``L\tau"$) is the heating power behind the shock that drives the explosion.  Conversely, the lowest compactness models that easily explode\footnote{e.g., 8.8-$M_{\odot}$ \citep{1984ApJ...277..791N} and $\sim$9.0 $-$ 9.5 $M_{\odot}$ at solar metallicity; 8.4 and 9.6 $M_{\odot}$ at very low or zero metallicity \citep[][and A. Heger, private communication]{wang_low_2024}} result in explosion energies of only $\sim$0.1 $-$ 0.15 Bethes \citep{burrows_2019,stockinger2020}. The physics behind this conclusion is also discussed in \citet{2021Natur.589...29B}, \citet{wang}, \citet{burrows_correlations_2024}, and \citet{burrows_40}.  Nevertheless, and importantly, compactness is but a crude substitute for the full mass density profile of the progenitor core at collapse, but it can capture the basic trend. A few-parameter characterization, perhaps including the outer mantle binding energy and/or multiple compactnesses defined at different interior masses, would be much preferred. 

\begin{figure*}   
    \centering
    \includegraphics[width=0.47\textwidth]{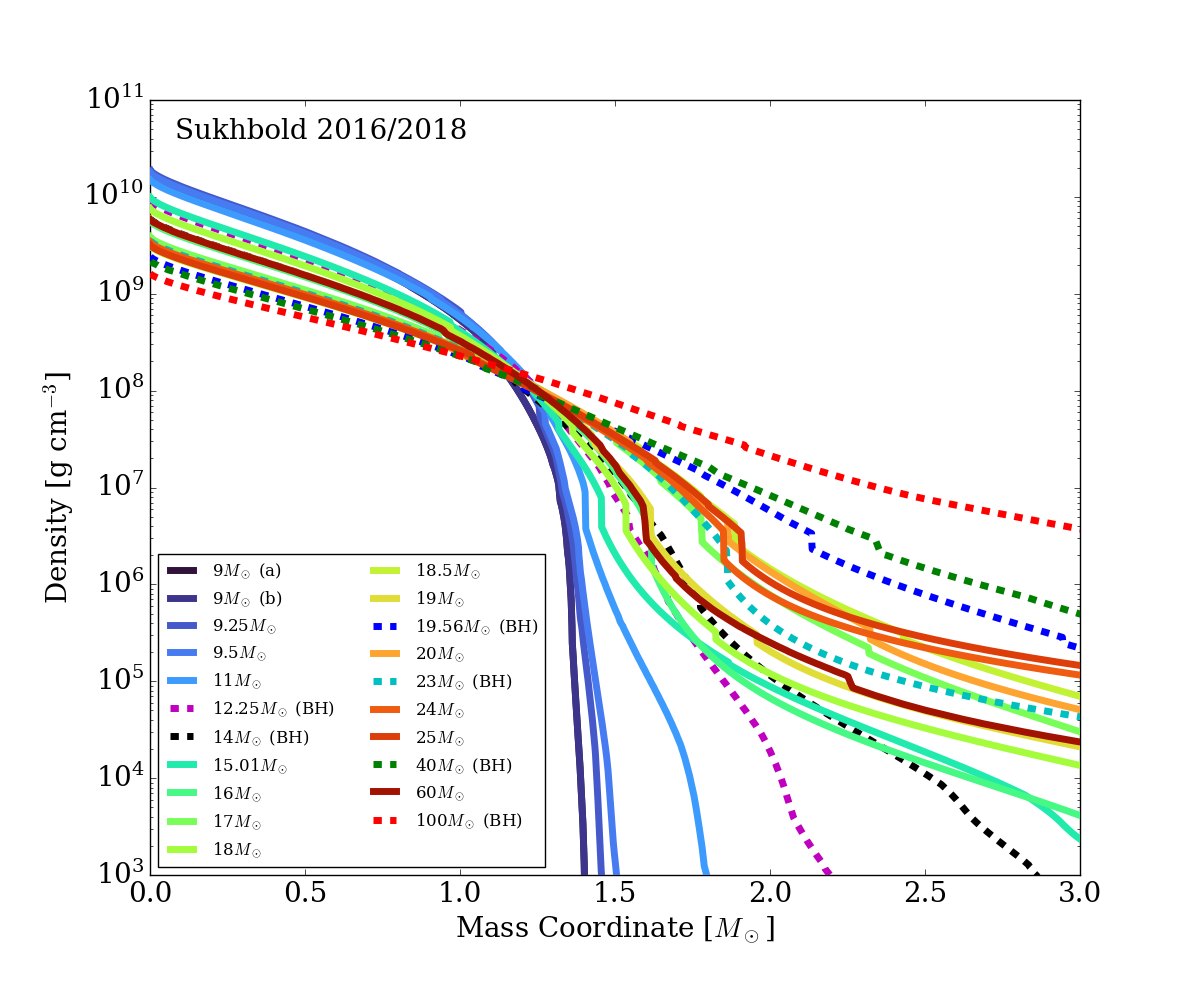}
    \includegraphics[width=0.47\textwidth]{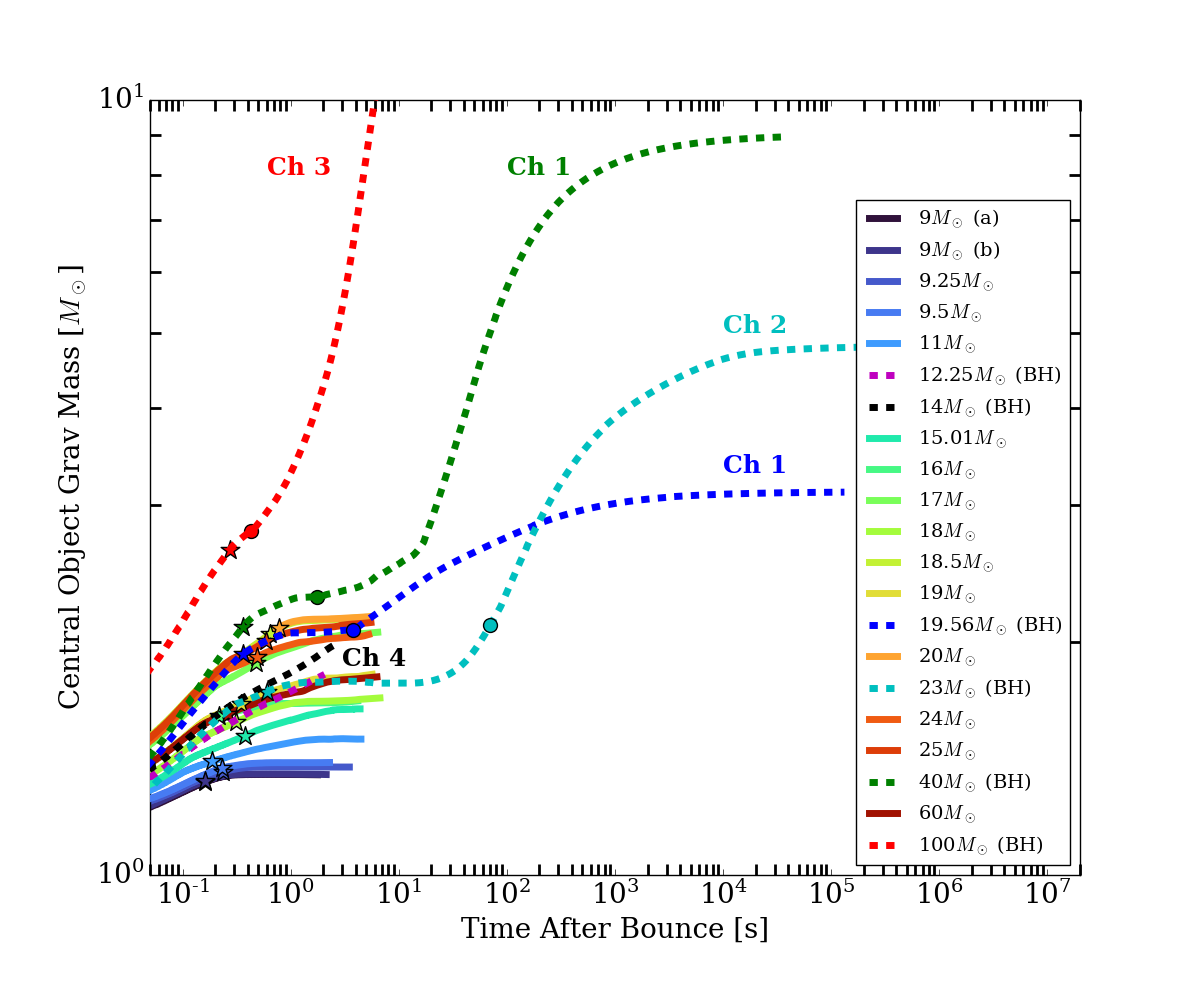}
    \caption{{\bf Left:} The progenitor mass density profile at collapse for many of our recent 3D CCSN simulations, colored by compactness ($\xi_{1.75}$) from low (violet) to high (red). The progenitor models were taken from \citet{swbj16} and \citet{sukhbold2018}. The dotted lines are for those models that formed, or will form, black holes. The others leave neutron stars. {\bf Right:} The evolution of the gravitational mass of the proto-neutron star (interior to 10$^{11}$ g cm$^{-3}$) with time after bounce for the same models depicted on the left. The coloring and the line types are the same as used on the left panel. The circle dots on the dotted lines indicate the time of black hole formation and the stars indicate the approximate time of explosion. See text for discussions of both these panels.}
    \label{fig:rho_full}      
\end{figure*}

\begin{figure*}
    \centering
    \includegraphics[width=0.48\textwidth]{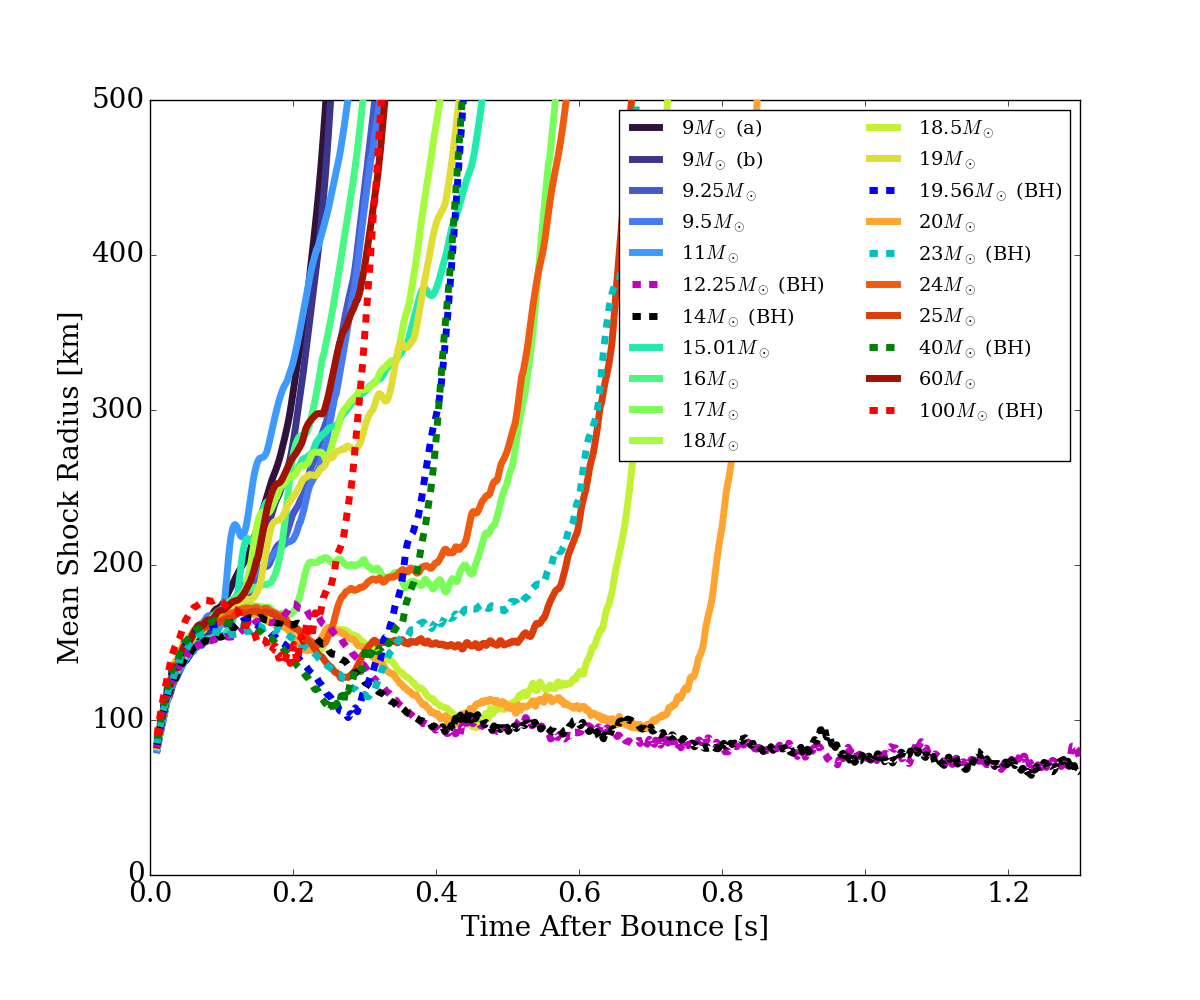}
    \includegraphics[width=0.48\textwidth]{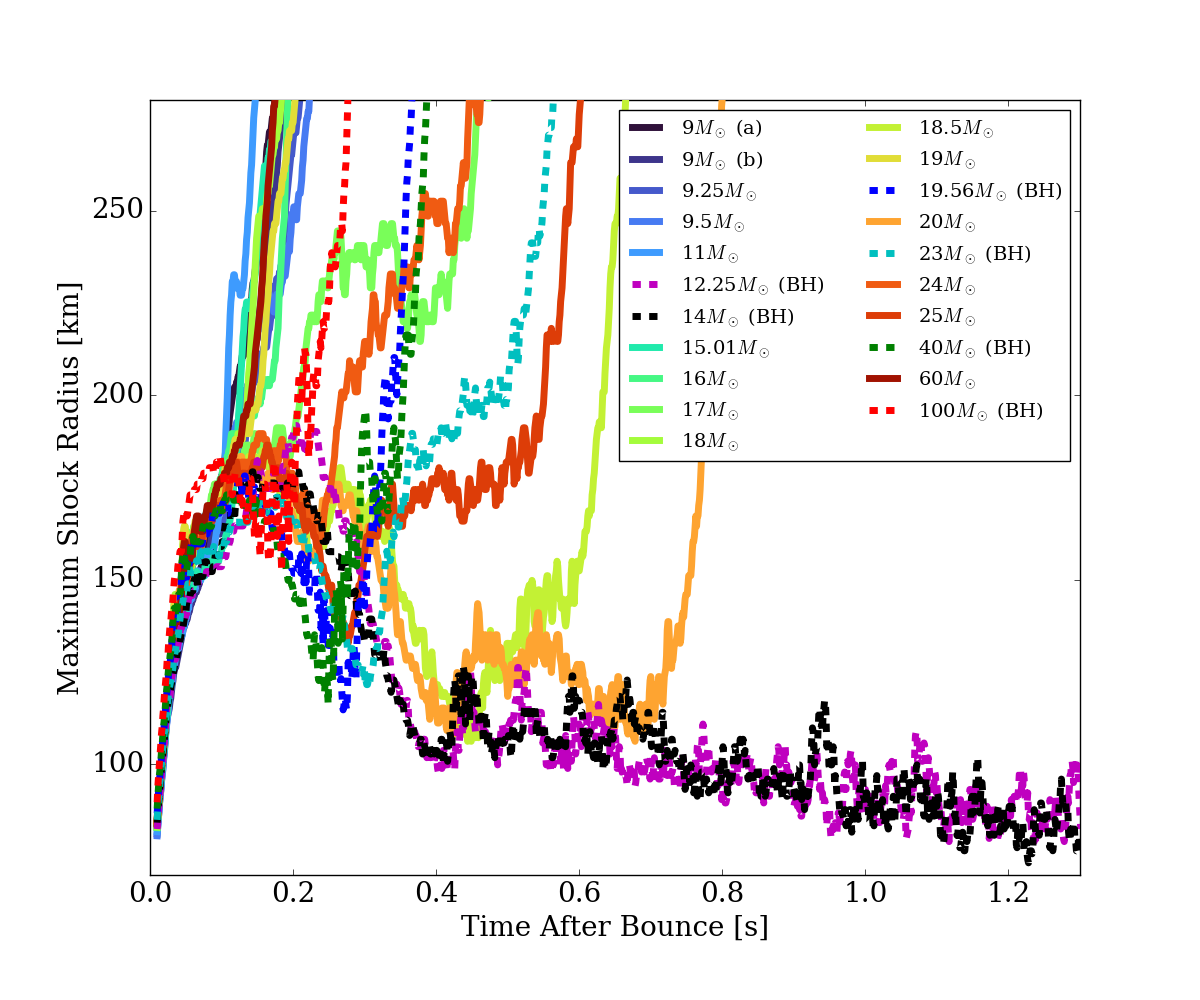}
    \caption{{\bf Left:} The mean shock radius versus time after core bounce for a collection of our detailed 3D CCSN models, colored by compactness ($\xi_{1.75}$). As the compactness increases the time to explosion first increases and then decreases for the highest compactnesses and associated mass accretion rates ($\dot{M}$). The dotted curves indicate those models that eventually form black holes via one of the channels discussed in this paper.  {\bf Right:} The same as on the left panel, but zoomed in and for the maximum shock radius versus time after bounce. The fluctuations are due to the spiral SASI (with frequency near $\sim$100 Hz) that when is appears is more manifest at the higher compactnesses. Note that for the highest compactness models (red), the onset of the spiral SASI is earliest and appears for larger values of the shock radius. The spiral SASI is all but absent for the lowest compactness models. See text for discussions.}
    \label{fig:rs}
\end{figure*}

\begin{figure*}
    \centering
    \includegraphics[width=0.90\textwidth]{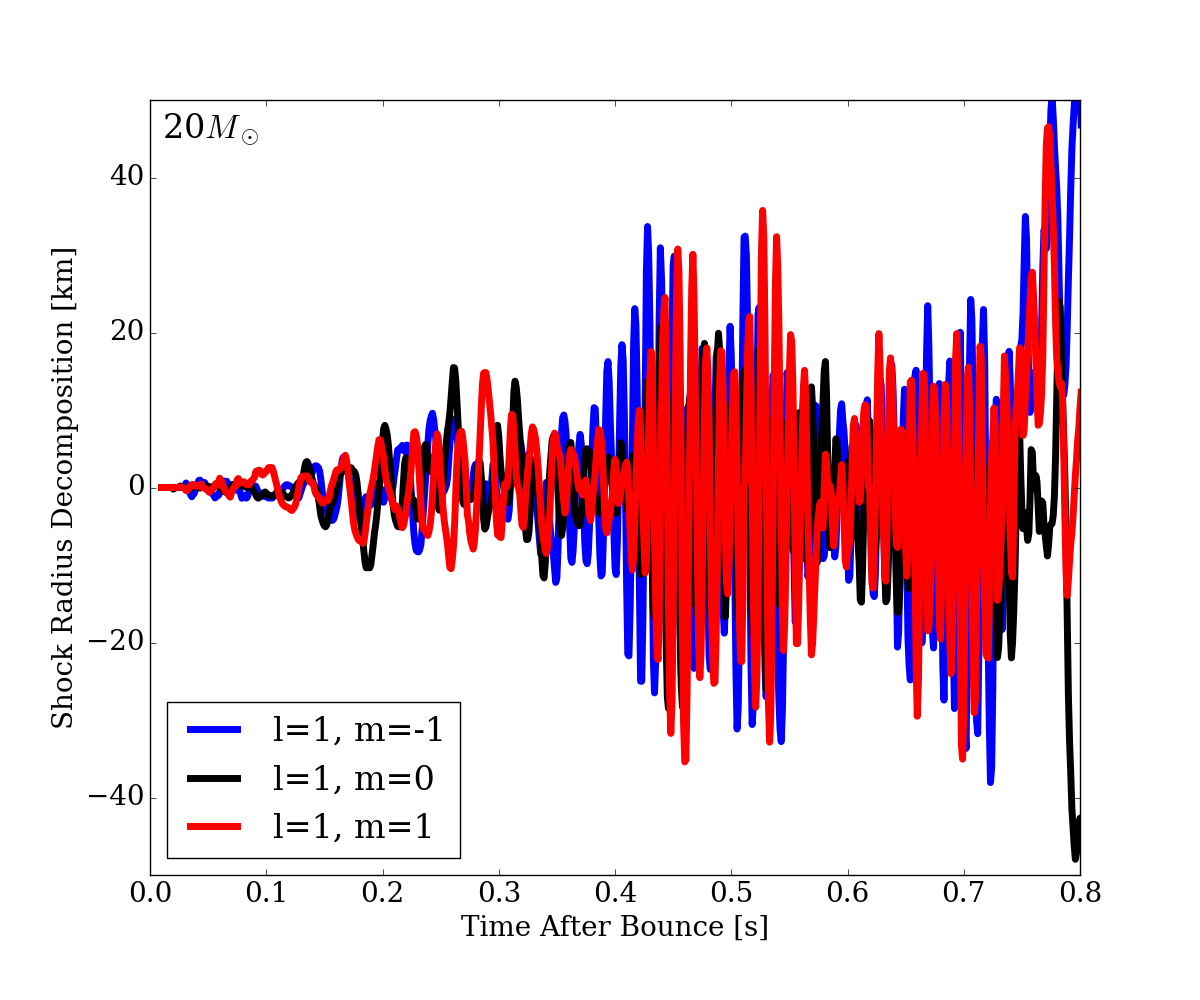}
    \caption{This figure depicts the three dipole components of the spherical harmonic decomposition of the shock radius as a function of time for the 20 $M_\odot$ model. The $\sim$100 Hz spiral-SASI mode becomes significant at around 0.4 seconds post bounce and lasts for about 0.3 seconds until the model explodes at about 0.7 seconds post bounce (See Figure \ref{fig:rs}).}
    \label{fig:dipole}
\end{figure*}

The left panel of Figure \ref{fig:rho_full} depicts the wide range of mass density profiles from the \citet{swbj16} and \citet{sukhbold2018} progenitor suite that we used to simulate CCSNe in 3D to late times using F{\sc{ornax}}. The models are colored by compactness at 1.75 $M_{\odot}$ ($\xi_{1.75}$). These profiles determine in full the accretion rate ($\dot{M}(t)$) history and, hence, the evolution and outcome for initially non-rotating stars. One of those outcomes is the final mass of the residue, depicted for our collection of 3D models in the right panel of Figure \ref{fig:rho_full}. Note that this figure demonstrates the tight relationship between residual mass (neutron star or black hole) and compactness \citep{burrows_correlations_2024}. 

Second, the asymptotic state of explosion energy, nucleosynthesis, recoil kick speeds, etc. is not achieved for most models before several seconds after bounce.  In addition, budgeting in the binding energy of off-grid stellar matter, the energy of the material exterior to the PNS is {\it negative} at the time of shock revival. Constant and continued neutrino heating after the onset of explosion is required to achieve not only positive energies (necessary for ejection to infinity), but to determine the final explosion energy \citep{2015MNRAS.453..287M,burrows_2020,burrows_correlations_2024}. Except in the case of the very lowest compactness progenitors, calculations that stop $\sim$0.3$-$$\sim$1.0 seconds after bounce, which is the majority of published 3D CCSN simulations, can say nothing about any of the final state observables, except perhaps the residual neutron star mass. In particular, the calculation and use of a so-called ``diagnostic" explosion energy that does not include the outer binding energy is fundamentally misleading. 

Third, the standing accretion shock instability (``SASI") \citep{blondin2003} is a vortical-acoustic \citep{foglizzo2002} instability of the shock wave shape that can arise when the shock radius recedes or when the mass accretion rate is large. An approximate condition for this is a small ``Foglizzo number" \citep{foglizzo:07}\footnote{The Foglizzo number is proportional to the ratio of the advection time between the shock and PNS surface and the convective overturn time, itself inversely proportional to the Brunt frequency. When the shock radius is small and/or when the mass accretion rate through the shock is large, this condition can be met.}. For most low-mass progenitors that explode prior to the significant recession of the shock radius, it does not appear and is completely overwhelmed by neutrino driven convection \citep{burrows2012,buellet2023}.  In previous 2D simulations, the ``sloshing" of the shock along the symmetry axis was often mistaken for the SASI, even though it was neutrino-driven convection. In 3D simulations, such axial sloshing is rarely seen. However, 1) for higher compactnesses and mass accretion rates prior to explosion; 2) at a few hundreds of milliseconds after bounce; and 3) generally for the higher mass progenitors, the mean shock radius recedes (even before explosion) and can settle from $\sim$175 kilometers to $\sim$100$-$125 kilometers \citep{burrows_2020}. At this point, whether or not the model explodes, the Foglizzo condition is met and a {\it spiral} SASI \citep{blondin_shaw} emerges. The emergence of such a mode was also noted by \citet{andresen2017} for a short interval around the time their 25-$M_{\odot}$ model started to explode. This has not an axial motion, but a quasi-periodic rotary motion with a period near $\sim$10 milliseconds, set approximately by the advection time between the shock and the inner PNS core \citep{foglizzo:07}. This mode modulates the shock dipole direction, the solid-angle-integrated neutrino emissions, and the gravitational-wave signal \citep{andresen2017,vartanyan2023}. It is seen in all our 3D simulations that do not explode early. It is also partially responsible for the roughly monotonic-with-compactness dependence seen in the degree of ejecta asymmetry \citep{burrows_correlations_2024}. All our 3D higher-compactness models that either explode and form neutron stars, explode and form black holes, or do not explode and later form black holes more quiescently manifest this late-time spiral SASI mode. Specifically, it is seen in all our black hole formation models in the CCSN context.  Figure \ref{fig:rs} depicts the evolution after bounce of both the mean (left) and maximum (right) shock radius with time for many of our 3D models. The cluster of higher compactness models (orange) are most delayed and clearly manifest the $\sim$100 Hz spiral SASI wobble during the launch phase. \adam{This spiral SASI behavior is more clearly seen in Figure \ref{fig:dipole}, which depicts the temporal variation of the dipole components of the shock surface for the 20 $M_{\odot}$ model.} 

One then asks: what is the significance of this compactness/spiral-SASI correlation? At low compactnesses (generally associated with the lowest mass progenitors), turbulence-aided, neutrino-driven explosions emerge early and easily. The accretion component of the neutrino luminosity is minimal for these and the explosion energies are correspondingly low. For these the spiral SASI seems to play no role in the explosion. 

As the mass density profiles shallow and the compactness and $\dot{M}$ rise, both the post-shock ram pressures and the accretion component of the driving neutrino luminosities increase. At intermediate compactnesses (perhaps $\xi_{1.75}\sim$0.3$-$0.5), one encounters using the \citet{swbj16} progenitor model suite the most difficult ``explodability" realm \citep{2020ApJ...890..127C,burrows_2020,burrows_correlations_2024,burrows_40,sykes_2024b}. The associated increase in the accretion luminosity (and in neutrino heating power behind the shock) seems not sufficient to counter the greater ram, unless the silicon/oxygen interface has a steep enough jump. Many of the \citet{swbj16} and \citet{sukhbold2018} models in this intermediate compactness regime (from $\sim$12 to 15 $M_{\odot}$) don't seem to have an adequately steep density jump. Figure \ref{fig:rs} depicts this behavior for our 12.25 and 14-$M_{\odot}$ models and we discuss this modality of black hole formation (Channel 4) in \S\ref{silent}. When it is steep enough, upon the accretion of such a jump the ram pressure would immediately decrease, while the neutrino luminosity driving the heating in the gain region \citep{wilson1985,1985ApJ...295...14B} behind the shock is maintained. This frequently kicks such a model into explosion.  

A bit higher in compactness, the shock still recedes and the spiral SASI still emerges. However, for the larger associated accretion luminosities and due to the slight growth in the gain region itself that follows the emergence of the spiral SASI, a vigorous explosion ensues. The longer delay of perhaps a few hundred milliseconds after bounce results in more vigorous neutrino-driven convection and a larger dipole asymmetry in the shock surface. The result is a more asymmetrical explosion, predominantly in the approximate direction of least confining ram.  In fact, as compactness further increases, we see higher explosion energies, with a roughly monotonic relation between the two \citep{burrows_correlations_2024}.
Concomitantly, there is a rough correlation of global asymmetry with explosion energy, though with exceptions that may be due to natural stochasticity in chaotic and turbulent flow and the fact that a single compactness alone does not fully capture the dynamical evolution. Importantly, perspectives that posit the primacy of low compactness as an explodability condition would be blind to this behavior.

In 3D models and for modest to high compactness you see simultaneous explosion (mostly in one direction) and accretion (mostly in others) \citep{vartanyan2018b}. This breaking of spherical symmetry and the consequent simultaneous explosion/accretion phenomenon can't happen, nor can it be adequately explored theoretically, in 1D. The supernova shock does emerge in all directions, but in some directions, due to its weaker strength in those directions, it initially fails to reverse the sign of the radial component of the infalling plumes. This early post-explosion infall is the origin of the continuing accretion onto the PNS core that provides the extra accretion luminosity component that helps to drive the early explosion, while simultaneously increasing the PNS mass. At the same time, in the other directions matter is indeed ejected.   

\begin{figure}    
    \centering
    \includegraphics[width=0.80\textwidth]{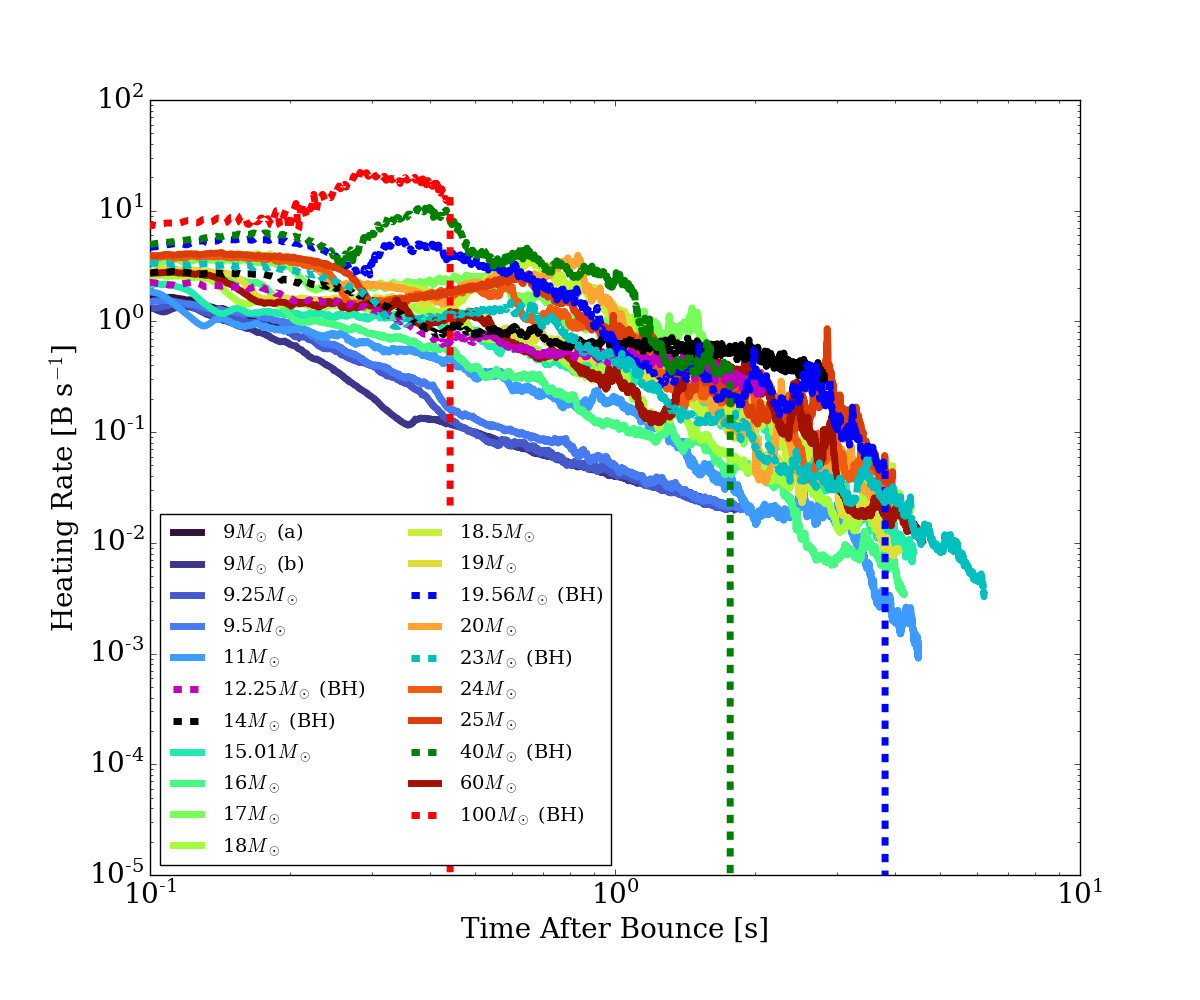}
    \caption{The power deposition (heating) rate (in Bethes per second) due to neutrino absorption in the gain region between the shock and the inner core as a function of log$_{10}$ time after bounce for a large collection of our recent 3D CCSN models, colored by compactness ($\xi_{1.75}$) from violet to red. This heating rate is what would drive explosion. The systematics of the heating rate with compactness (and the associated $\dot{M}$) is quite clearly displayed. The dotted models are for those that eventually form black holes according to our current calculations. The abrupt drops to ``zero" clearly in evidence for the black-hole formers at the highest compactnesses happen at the corresponding times of black hole formation.}
    \label{fig:Qdot}
\end{figure}

\begin{deluxetable*}{cccccccc}
\tablecolumns{8}
\tablewidth{0pt}
\label{table1}

\begin{minipage}{\textwidth}
  \centering
  \textbf{Basic Model Data} \\  
\end{minipage}

\tablehead{\colhead{Progenitor} & \colhead{$\xi$(1.75,2.5)}& \colhead{F{\sc{ornax}}(FLASH) Time (s)} & \colhead{$M_G$ ($M_{\odot}$)} & \colhead{Energy (Bethes)} & \colhead{Kick (km s$^{-1}$)}& \colhead{$^{56}$Ni ($M_{\odot}$)} & \colhead{Ejected Mass}  }
\startdata
9a      & (6.7, 3.8) $\times 10^{-5}$ & 1.775 & 1.237 & 0.111 & 120.7 & 0.168$\times$10$^{-2}$ & 7.4 \\
9b      & (6.7, 3.8) $\times10^{-5}$  & 2.14 & 1.238 & 0.095 & 78.6    & 0.612$\times$10$^{-2}$ & 7.4    \\
9.25    & (2.5, 0.047)$\times10^{-3}$ & 3.532 & 1.263 & 0.124 & 140.1    & 1.04$\times$10$^{-2}$  & 7.6    \\
9.5     & (8.5, 0.062)$\times10^{-3}$ & 2.375 & 1.278 & 0.143 & 208.6    & 1.47$\times$10$^{-2}$  & 7.8    \\
11      & 0.12, 0.0076                & 4.492 & 1.361 & 0.326 & 699.4    & 2.92$\times$10$^{-2}$  & 9.2    \\      
{\bf 12.25}   & {\bf 0.34, 0.030}     & {\bf 2.090} &  {\bf $\sim$11.1}    & {\bf $-$} & {\bf 6.5}  &  {\bf $-$}  &  {\bf $?$ }   \\
{\bf 14}      & {\bf 0.48, 0.12}      & {\bf 2.824} & {\bf $\sim$12.1}    & {\bf $-$} & {\bf 7.0} & {\bf $-$}   & {\bf $?$}    \\
15.01   & 0.29, 0.13                  & 4.5 & 1.474 & 0.352 & 173.9    & 5.42$\times$10$^{-2}$  &10.9     \\
16      & 0.35, 0.12                  & 4.184 & 1.505 & 0.463 & 468.0    & 6.06$\times$10$^{-2}$  &11.4    \\
17      & 0.74, 0.24                  & 6.390 & 1.794 & 1.266 & 910.8    & 9.99$\times$10$^{-2}$  &11.6     \\
18      & 0.37, 0.15                  & 8.508 & 1.516 & 0.600 & 735.7    & 10.3$\times$10$^{-2}$  &12.6     \\
18.5    & 0.80, 0.31                  & 6.359 & 1.862 & 1.254 & 821.7    & 13.8$\times$10$^{-2}$  &12.5     \\
19      & 0.48, 0.17                  & 7.004 & 1.612 & 0.561 & 525.7    & 7.73$\times$10$^{-2}$  &12.6     \\
{\bf 19.56}   & {\bf 0.85, 0.45}      & {\bf 3.890 (1.3$\times$10$^{5}$)} & {\bf 3.12 }& {\bf $\sim$2.5} & {\bf $\sim$1300}   & {\bf 20.0$\times$10$^{-2}$}  & {\bf $\sim$10.5 }    \\
20      & 0.79, 0.28                  & 6.337 & 1.865 & 0.983 & 591.6    & 9.94$\times$10$^{-2}$  &12.9     \\
{\bf 23}      & {\bf 0.74, 0.21}      & {\bf 6.228 ($\sim$3$\times$10$^{5}$)} & {\bf 4.9} & {\bf $\sim$0.46} & {\bf $\sim$90}    & {\bf 3.9$\times$10$^{-2}$}  & {\bf $\sim$9.9}    \\
24      & 0.77, 0.27                  & 6.293 & 1.784 & 0.883 & 1069.0    & 12.5$\times$10$^{-2}$  &12.6     \\
25      & 0.80, 0.30                  & 6.324 & 1.838 & 1.301 & 677.4    & 16.8$\times$10$^{-2}$  &13.8     \\
{\bf 40}      & {\bf 0.87, 0.54}      & {\bf 1.760/21 (38000)} & {\bf $\sim$9.0} & {\bf $\sim$1.75}  & {\bf $\sim$550}  & {\bf 11.4$\times$10$^{-2}$}  & {\bf $\sim$6.18}\\
60      & 0.44, 0.17                  & 7.899 & 1.602  &0.688 & 926.3    & 10.6$\times$10$^{-2}$  &5.5     \\ 
{\bf 100}     & {\bf 1.02, 0.81}      & {\bf 0.442 (100)} & {\bf $\sim$37} & {\bf $-$} & {\bf $\sim$0.0}    & {\bf 0.0}      & {\bf }    \\ \hline   
\enddata
\caption{Summary of some basic physical and observational quantities of the 3D F{\sc{ornax}} model simulations described here and in \citet{burrows_correlations_2024}. Those simulations that leave black holes are put in bold. $M_G$ is the gravitational mass left behind (whether it is a black hole or neutron star), ``Energy" is the explosion energy, ``Kick" is the net recoil speed due to the inner CCSN dynamics and net neutrino emission asymmetry, ``$^{56}$Ni" is the $^{56}$Ni mass ejected to infinity that does not fall back, ``Ejected Mass" is our total supernova ejecta mass in $M_{\odot}$ using the single solar-metallicity models of \citet{swbj16} and \citet{sukhbold2018}, and $\xi(1.75,2.5)$ is the compactness at both 1.75 and 2.5 $M_{\odot}$ interior masses for the unstable progenitor core. The F{\sc{ornax}}(FLASH) times are the physical times after bounce in seconds simulated by each code. For the 100-$M_{\odot}$ model we leave the energy and ejected mass blank since we did not calculate these values (see S. Woosley, in preparation).{The final kick velocities of the black hole formers are estimated by dividing the central object's momentum at mapping from the F{\sc{ornax}} to the FLASH phases with the final central object gravitational mass.}}
\end{deluxetable*}

Notably, as compactness and mass accretion rate increase further, the tendency for the spiral SASI to arise and the PNS mantle to explode asymmetrically (and with simultaneous accretion and explosion) does not abate.  As a result, another related consideration simultaneously comes into play. The higher $\dot{M}$s onto the core naturally fatten the residue at a progressively greater rate with increasing compactness. The question then arises: does this push the PNS mass above the thermally-corrected maximum mass of a neutron star?  Does a black hole form early?  And what of the explosion?  The majority of our solar-metallicity \citet{swbj16}- and \citet{sukhbold2018}-derived 3D models from 16 to 25 $M_{\odot}$ with modest to high compactnesses and infall $\dot{M}$s eventually cease accretion and have explosion energies large enough to eject the rest of the star. For most of these models, the infall is eventually reversed, in part by the neutrino-driven winds that later emerge \citep{wang_wind}, and the core mass freezes in the stable neutron star range (see right panel of Figure \ref{fig:rho_full}). However, at the highest compactnesses for our current collection of 3D models (the 19.56-, 40-, and 100-$M_{\odot}$ progenitors), the accretion rate onto the PNS core is so high that a black hole indeed forms and neutrino heating ceases.  Before the general-relativistic instability sets in, the neutrino heating rate in the gain region was for these models the highest observed. Figure \ref{fig:Qdot} demonstrates this quite starkly. For the 19.56- and 40-$M_{\odot}$ models, the integral over time of this heating rate was sufficiently high to launch vigorous explosions. We discuss this modality of stellar-mass black hole formation (Channel 1) in \S\ref{40_19.56_explosion}. However, for our highest compactness progenitor (the 100-$M_{\odot}$/tenth-solar PPISN model) for which the driving heating rate was higher still, the corresponding $\dot{M}$ onto the core was higher as well. The result in this case was black hole formation at such an early time ($\sim$440 milliseconds after bounce) that the time integral of the heating rate between launch and general-relativistic collapse was inadequate to provide an explosion energy sufficient to eject its even larger binding energy stellar envelope.  The net effect was a very asymmetrical explosion that within seconds reversed. The result was that the black hole grew eventually to $\sim$37 $M_{\odot}$. We discuss this modality of stellar mass black hole formation (Channel 3) in \S\ref{100_tenth}. Solar-metallicity models generally would have stellar winds during their pre-collapse quasi-static evolution sufficient to leave much lower total progenitor terminal masses (for the solar-metallicity \citet{swbj16} and \citet{sukhbold2018} stars this is $\sim$16 $M_{\odot}$ at maximum). 

Hence, there is a race at the highest compactnesses between mass accumulation of the core and the deposition of neutrino energy\footnote{This race is also dependent upon the nuclear equation of state.}. For the very highest compactnesses, generally more easily achieved at low metallicities and high progenitor masses, the supernova often loses this race.  The post-launch accretion rates and outer mantle binding energies are too great. Due to weaker stellar winds, for low-metallicity progenitors the corresponding final black hole masses should be higher \footnote{This expectation is consonant with the Gaia detection of the 33-$M_{\odot}$ BH3 with a low-metallicity companion \citep{gaia3_2024}.}, while at solar-metallicity they should be lower.  Indeed, for our highest-compactness solar-metallicity models the residual black hole masses span a range.  However, in all such cases the supernova is always (at least for the models we have explored) launched, and sometimes vigorously enough to result in a supernova explosion.  Indeed, our 19.56- and 40-$M_{\odot}$ models have the highest asymptotic supernova energies (including the stellar mantle binding energy!) we have yet simulated. 

There is a very important exception, which we find for our 23-$M_{\odot}$ model.  It has a low explosion energy ($\sim$0.46 Bethes) that is lower than the trend we generally find with compactness \citep{burrows_correlations_2024}. Its compactness, though high, is at a local minimum for $\xi_{2.5}$ (Table \ref{table1}).  Its mantle binding energy exterior to 2.0 $M_{\odot}$ is higher than we see for our binding energy/explosion energy correlation. From our F{\sc{ornax}} simulation it seemed that the PNS mass stopped increasing and there would be no late-time fallback. However, when we mapped the F{\sc{ornax}} model into FLASH and continued the explosion for many hours we witnessed significant late-time fallback of $\sim$3.2 $M_{\odot}$\footnote{As an aside, we observe that short-term (on second timescales) post-explosion infall accretion should be distinguished from what is often called ``fallback." The latter can occur on timescales much longer than a few seconds (many seconds to hours to days) and when originally imploding plumes reverse in velocity for a time to achieve positive radial speeds upon encountering the primary supernova shock wave, only eventually to fall back onto the core upon doing work on the exterior. In contrast, infall is very common in 3D core collapse models right after the ignition of an aspherical explosion, a context in which models generically experience simultaneous explosion and accretion (infall) \citep{2021Natur.589...29B,wang,vartanyan2018b}. The sign of its velocity is generally never reversed by the explosion shock. Such infall accretion onto the proto-neutron star powers a large fraction of the neutrino luminosity that drives the asymmetrical explosion in the first seconds after bounce and would be all but absent in spherical symmetry. It is a 3D effect. Fallback is likely on much longer timescales, may be less common than infall, and seems associated with weak explosions in highly bound outer envelopes. The distinction between the two is important, though they are frequently confused.}. On $\sim$day timescales, a black hole of mass $\sim$4.9 $M_{\odot}$ was formed, though a supernova explosion ensued with roughly the same launch energy and a final ejecta mass of $\sim$10 $M_{\odot}$. Since the fallback mass is a sensitive function of explosion energy and mantle density and binding energy profiles, and since in the chaotic, turbulent context of core-collapse supernova explosions there is bound to be some degree of stochasticity in explosion parameters, we surmise that a wide range of fallback and residual black hole masses could result via this channel. We discuss this formation channel (Channel 2) for stellar-mass black holes in \S\ref{bh_gap_weak}. We note that in broad outline this scenario was envisioned by \citet{chan2020} and may be a dominant modality for black hole formation leading to $\sim$4$-$12 $M_{\odot}$ black holes.


   
\section{Black Hole Formation Channels}
\label{channels}

We now proceed to discuss examples of each of the stellar-mass black hole formation modalities we observe. Note that a so-called ``direct collapse" to a black hole is impossible in the Chandrasekhar instability context $-$ one must always go through a proto-neutron star intermediary. The mass interior to the inevitable bounce shock ($\sim$1.2$-$1.5 $M_{\odot}$) is always initially far below the effective maximum mass of a neutron star. Since the PNS is out of sonic contact with the outer accreting matter, the initial PNS does not know that it might eventually fatten sufficiently to collapse to a black hole until the requisite additional matter has indeed been accreted through the shock. This takes time, whose duration is a function of the subsequent accretion rate (a function crudely of compactness) and the details of the complicated explosion/no-explosion dynamics.

The different primary stellar-mass black hole formation pathways we discuss in this paper are represented by four distinct channels. For each channel we provide example models: {\bf Channel 1:} asymmetric, high-compactness vigorously exploding solar-metallicity 40-$M_{\odot}$ and 19.56-$M_{\odot}$ models (\S\ref{40_19.56_explosion}) that within seconds of bounce form black holes, but later due to subsequent fallback may leave a spectrum of black hole masses; {\bf Channel 2:} a weakly exploding, mildly asymmetric, solar-metallicity 23-$M_{\odot}$ model (\S\ref{bh_gap_weak}) that experiences late-time fallback accretion and black hole formation on $\sim$1000-second timescales.; {\bf Channel 3:} a 100-$M_{\odot}$ tenth-solar model (\S\ref{100_tenth}) that experiences a PPISN and then an aborted highly-asymmetrical supernova explosion that initially sends out a shock wave to thousands of kilometers which weakens significantly into a sound wave. Within hundreds of milliseconds, its PNS collapses into a black hole and within tens of seconds that black hole fattens to $\sim$37 $M_{\odot}$; and {\bf Channel 4:} two models (12.25- and 14-$M_{\odot}$) (\S\ref{silent}) that never reignited their stalled shock wave and would accrete over minutes to hours timescales until ``quiescent" black hole formation.  This is the oft-assumed ``silent" channel. That this progenitor mass interval might evolve through this channel is quite unexpected and may be wrong.  However, using the \citet{sukhbold2018} progenitor suite, this is what we and others \citep{2020ApJ...890..127C,2021ApJ...914..140P,sykes_2024b} currently conclude.  We do not discuss the secondary black hole formation channels through accretion onto a neutron star from a companion star, the merger of two neutron stars, or the late-time developments of a Thorne-Zytkov object \citep{thorne1977}, all potentially important pathways, but outside the CCSN context.

\begin{figure}
    \centering
    \includegraphics[width=0.47\textwidth]{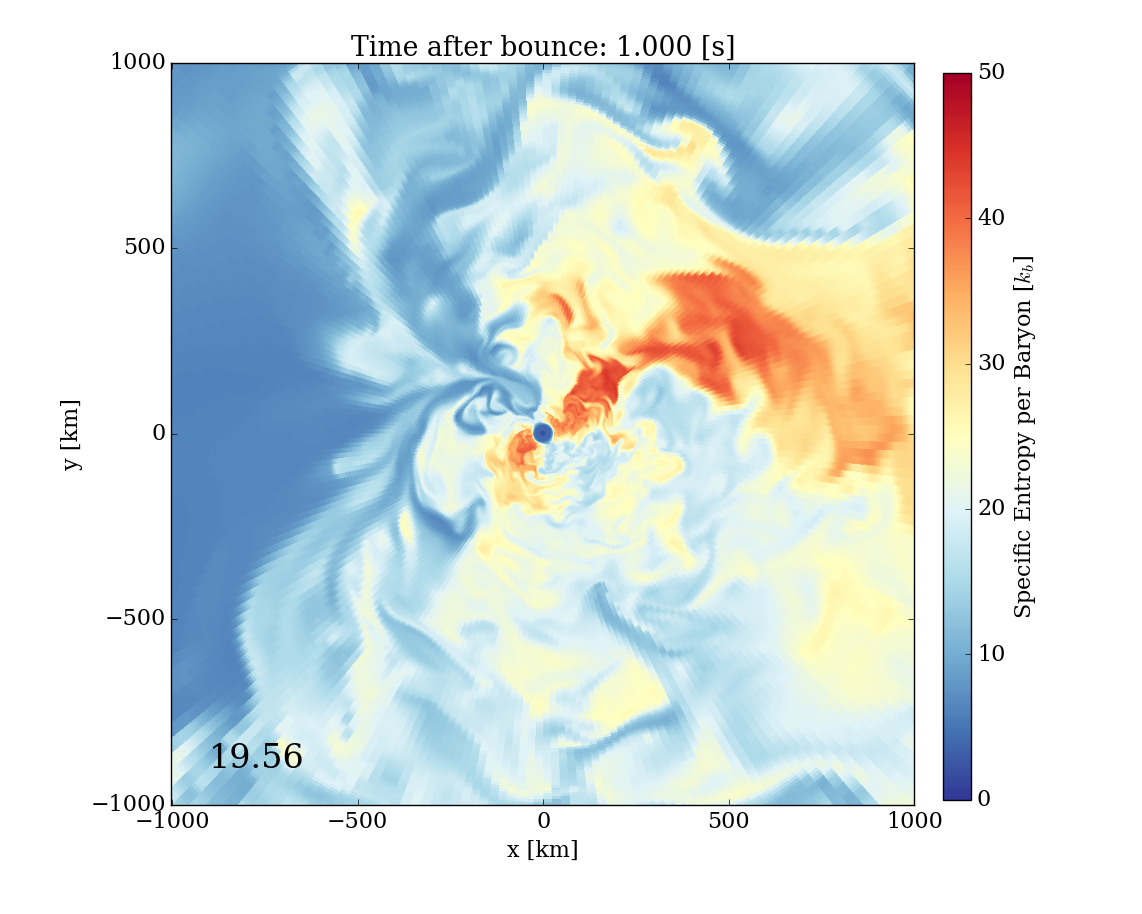}
    \includegraphics[width=0.47\textwidth]{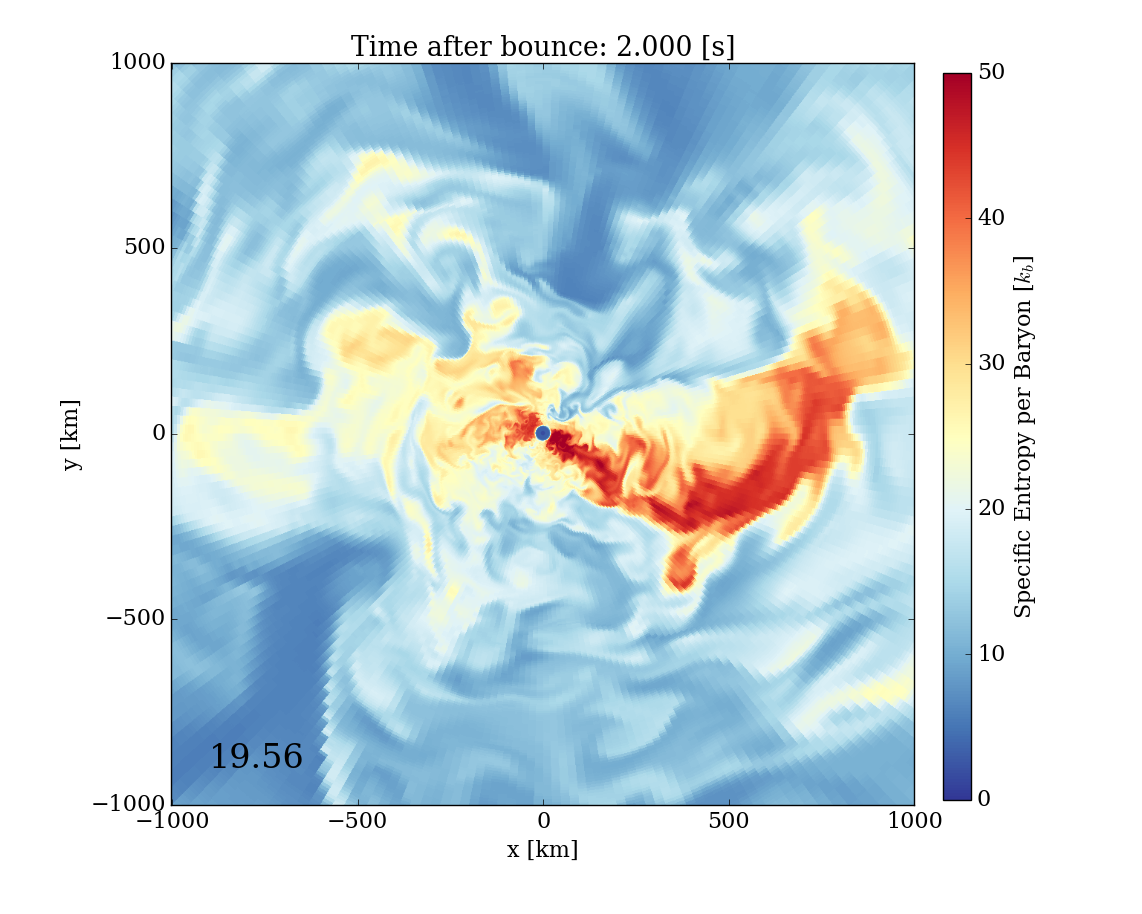}
    \includegraphics[width=0.47\textwidth]{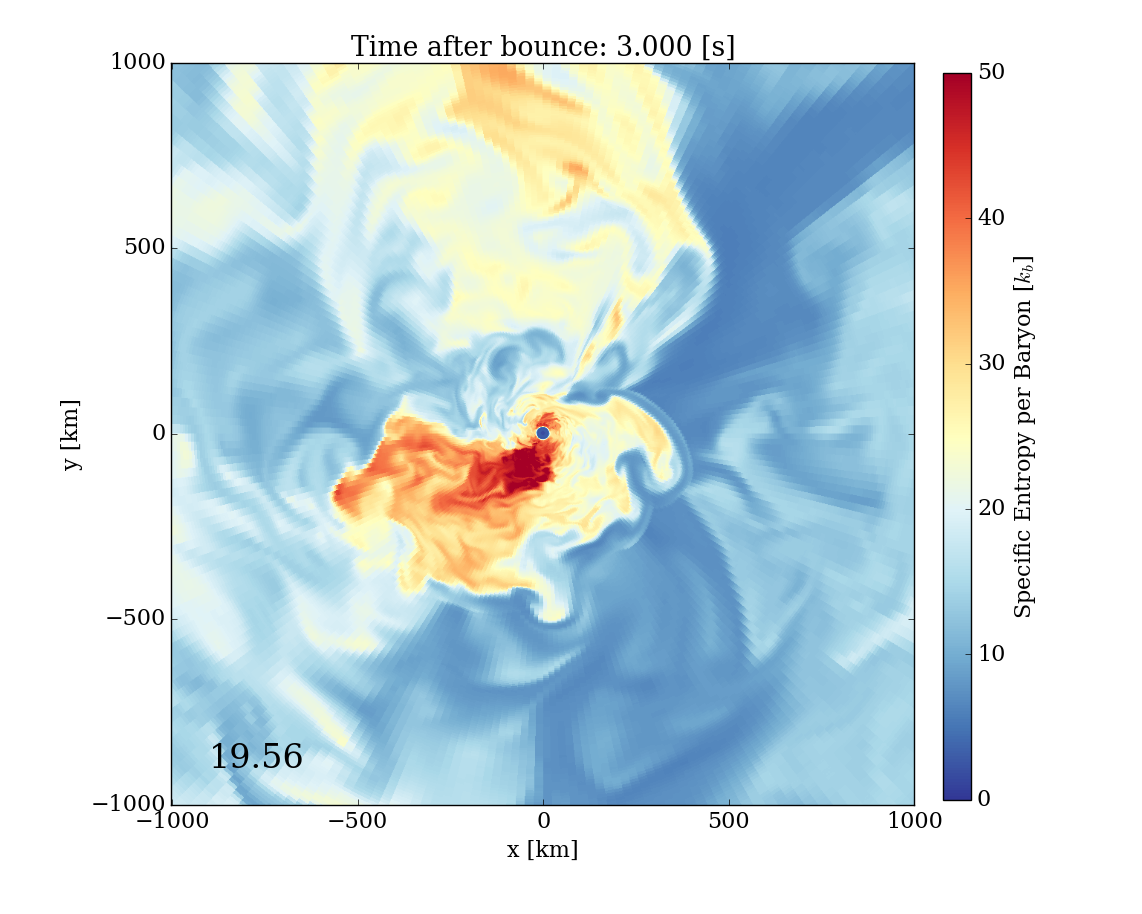}
    \includegraphics[width=0.47\textwidth]{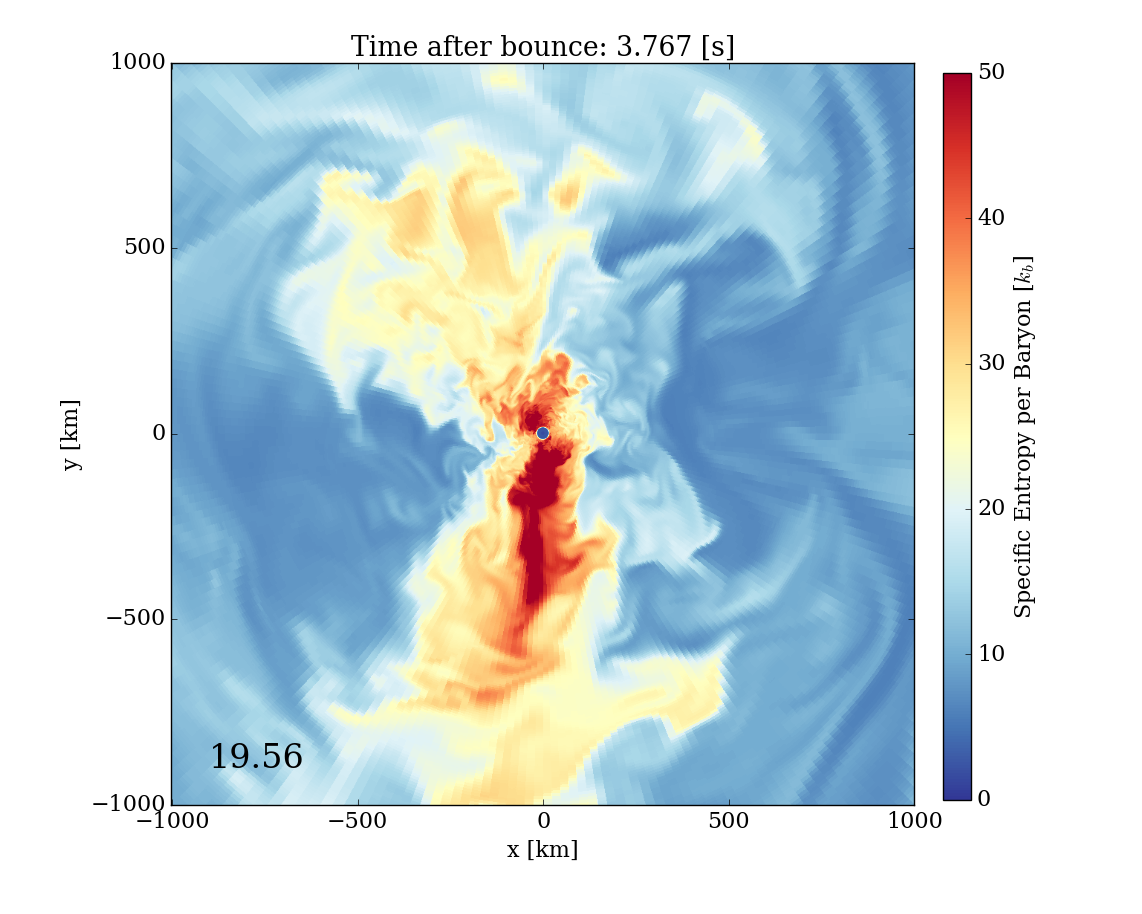}
    \caption{Snapshots of an x-y slice in entropy space (per baryon per Boltzmann's constant) of the inner region ($\pm$1000 km) of the explosion of the solar-metallicity 19.56-$M_{\odot}$ progenitor. These stills are taken at 1.0, 2.0, 3.0, and 3.767 seconds after bounce, the latter just before black hole formation. They reveal an asymmetric quasi-jetlike structure flapping about in response to the variations in the infalling plumes that are dancing over the PNS. The red regions are high-entropy neutrino-driven ejected, while the blue regions are mostly these lower-entropy (and more dense) infalling streams. Note the blue and red regions are ``complementary" $-$ the driven ejecta avoid the infalling fingers.  The degree of net asymmetry is correlated with the magnitude of the resultant recoil kick, with the corresponding neutrino emission anisotropy a small contributing factor. The 19.56-$M_{\odot}$ model explodes vigorously and asymmetrically and leaves behind a black hole. See text for discussion.}
    \label{fig:before_BH_19_1}
\end{figure}

\begin{figure}
    \centering
    \includegraphics[width=0.47\textwidth]{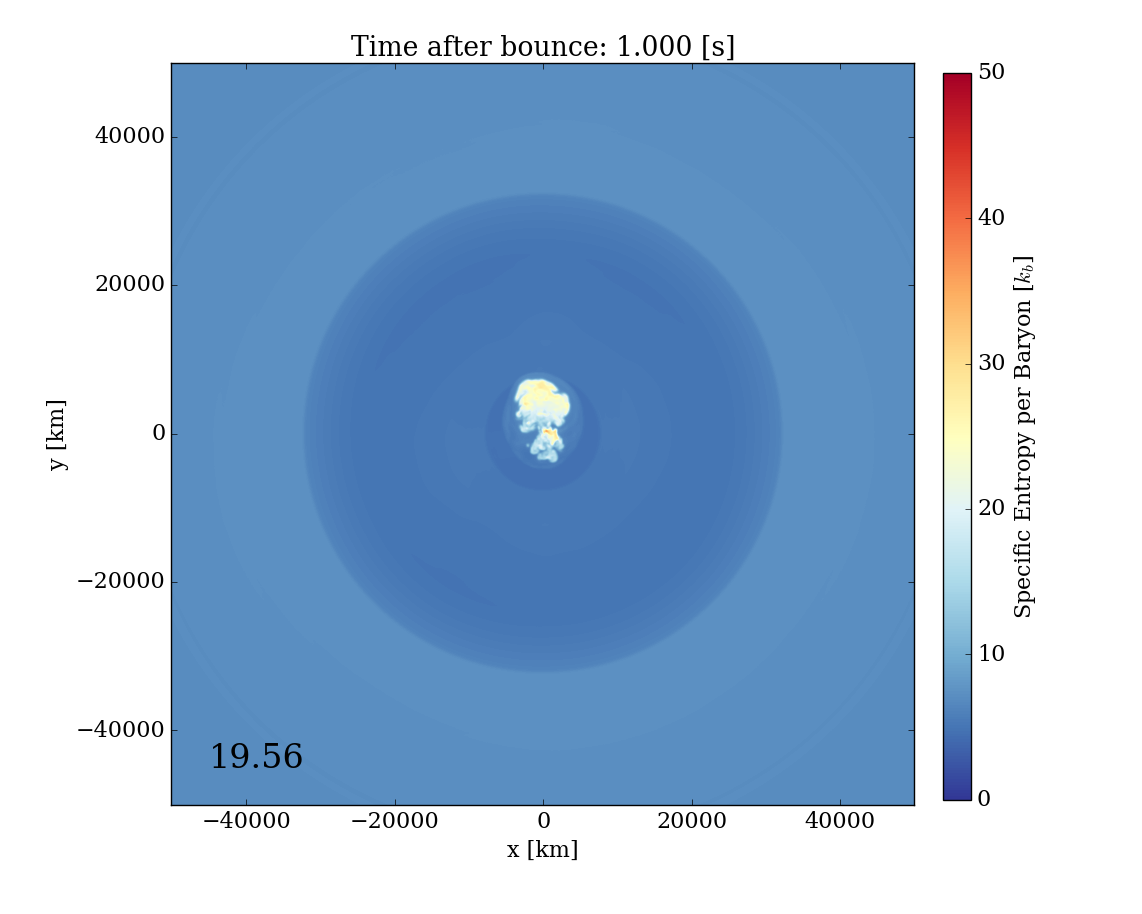}
    \includegraphics[width=0.47\textwidth]{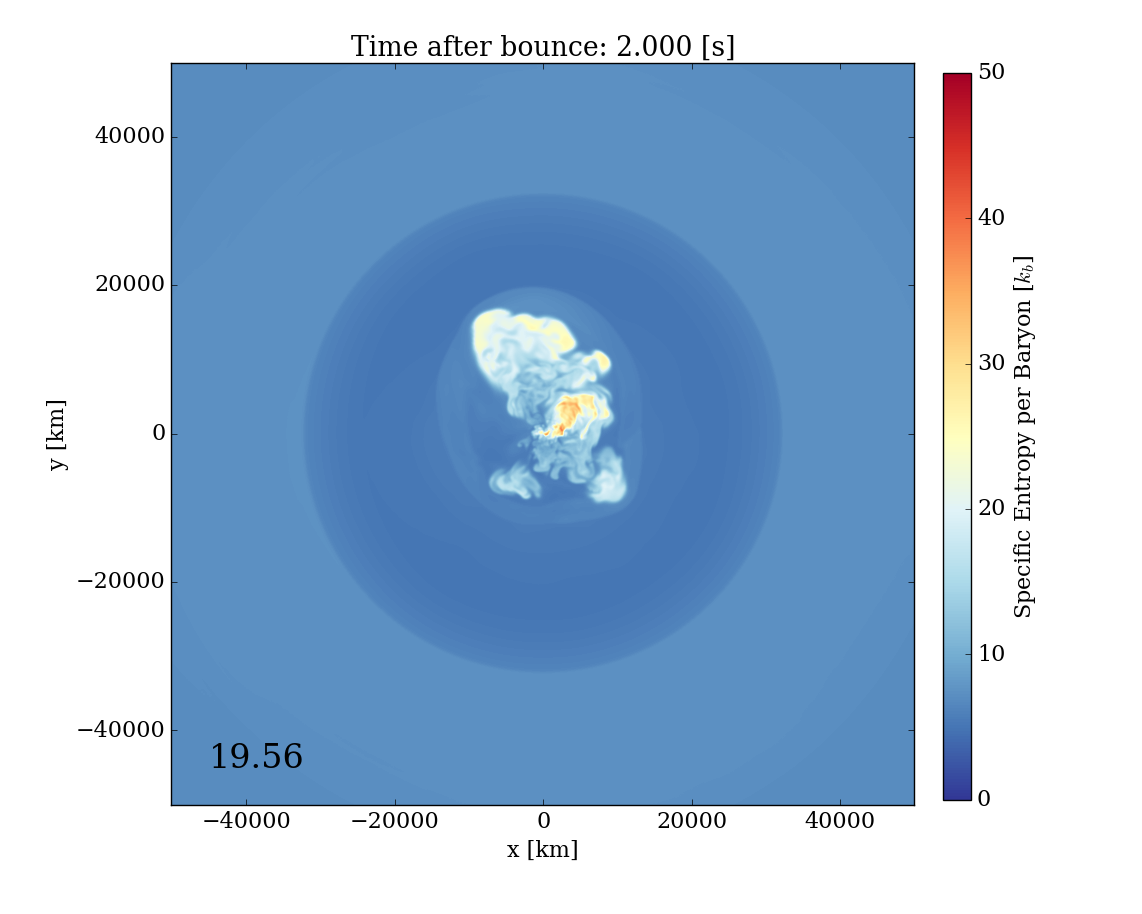}
    \includegraphics[width=0.47\textwidth]{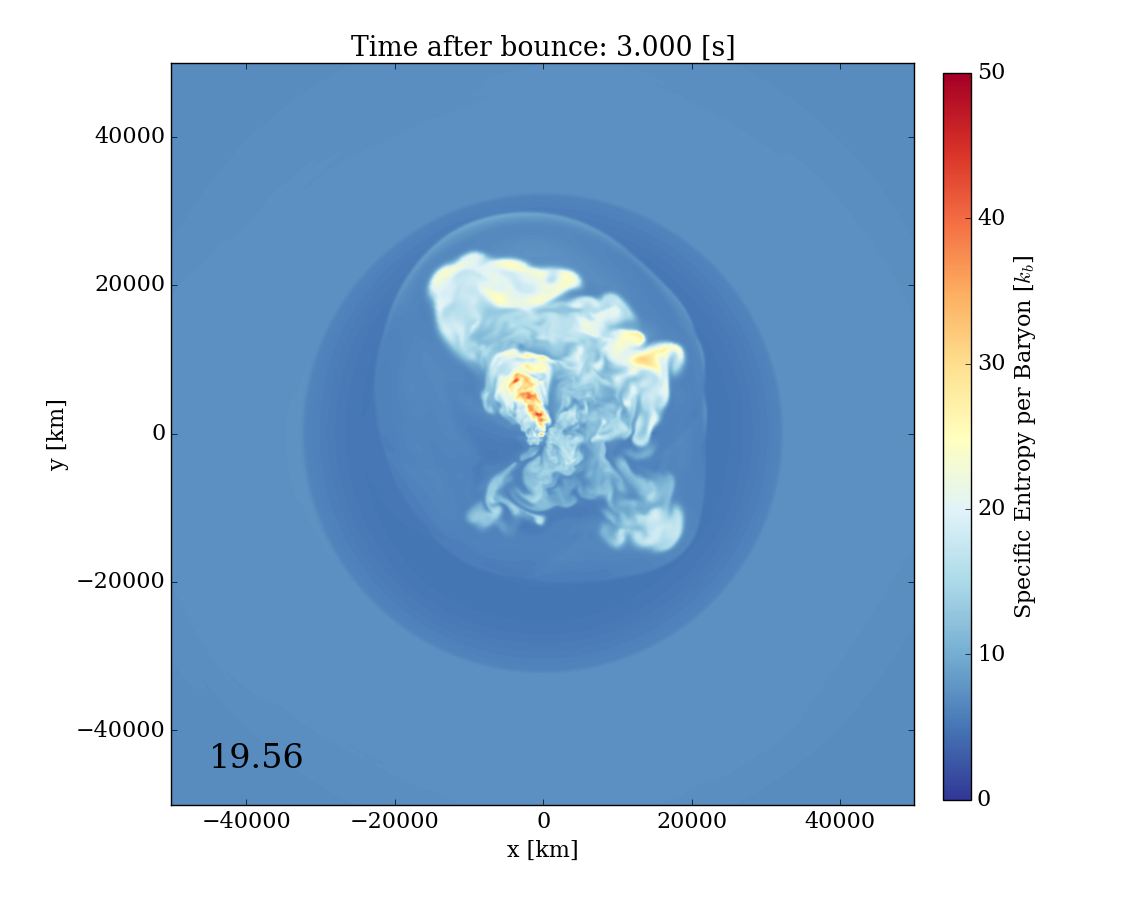}
    \includegraphics[width=0.47\textwidth]{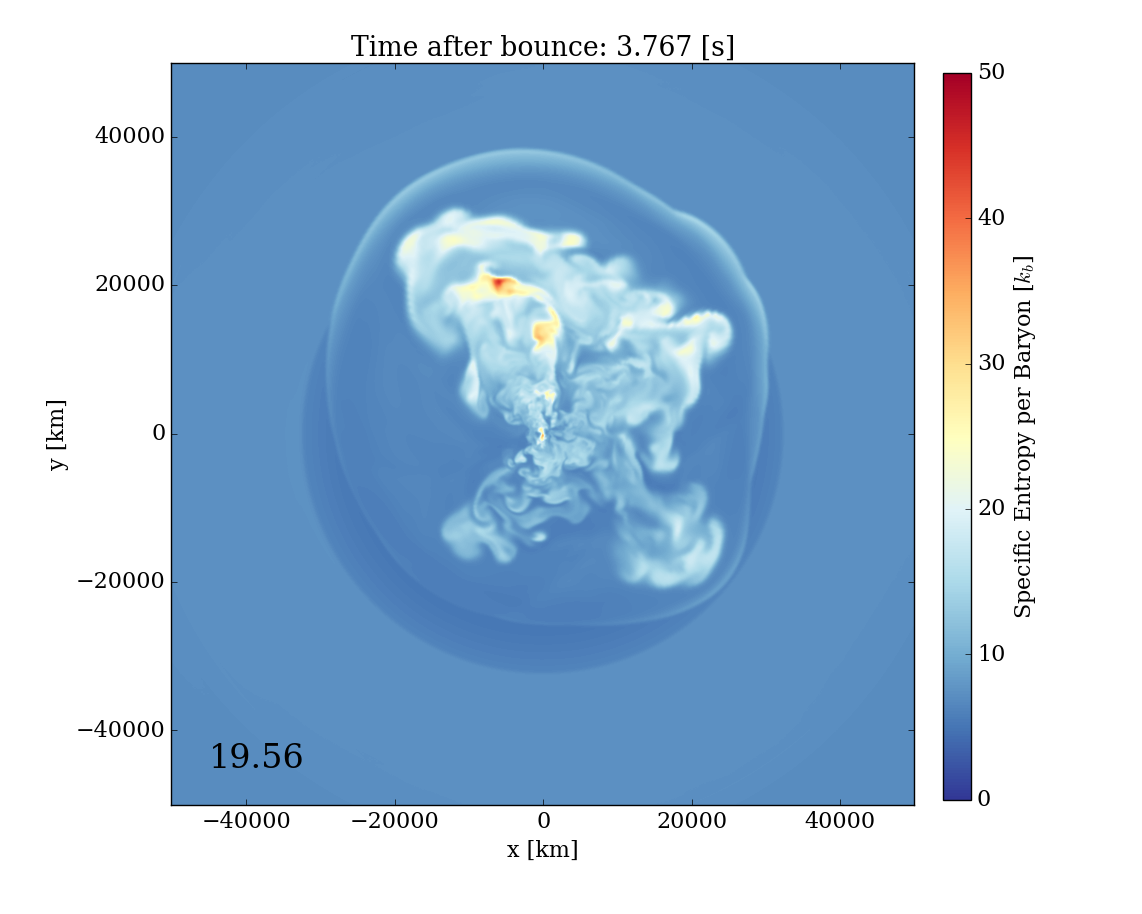}
    \caption{Sample snapshots of x-y slices in entropy space (per baryon per Boltzmann's constant) of the early ``F{\sc{ornax}}" phase of explosion of the solar-metallicity 19.56-$M_{\odot}$ model on a much larger scale than Figure \ref{fig:before_BH_19_1}. From top left to bottom right are stills at 1.0, 2.0, 3.0, and 3.767 seconds, respectively, after bounce. The black hole forms at $\sim$3.8 seconds. The blue-white veil is the blast wave. The light-blue to dark-blue transition marks the oxygen/carbon interface.  See text for a discussion.}
    \label{fig:before_BH_19}
\end{figure}

\begin{figure}
    \centering
    \hskip-0.0in\includegraphics[width=0.49\textwidth]{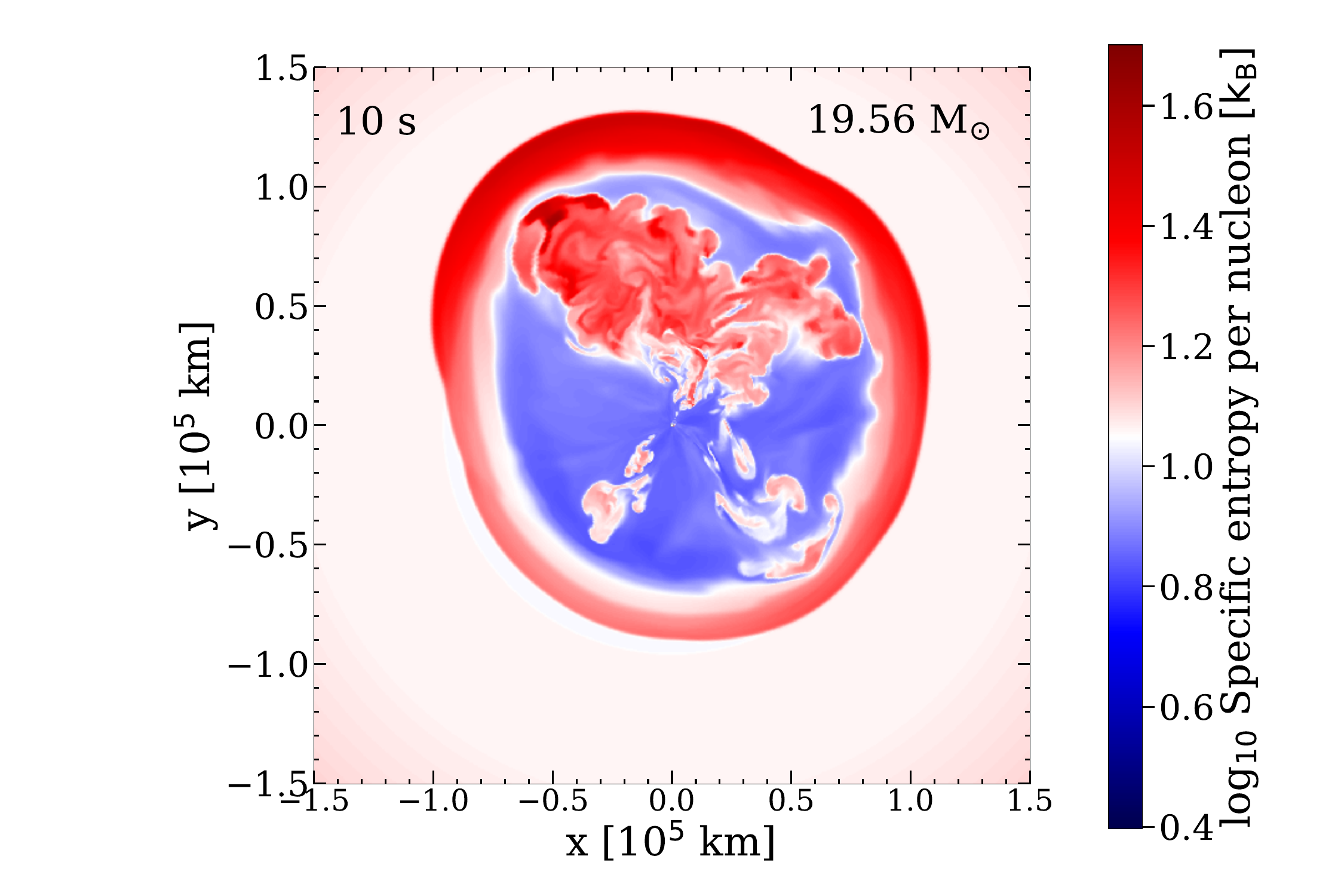}
    \hskip-0.0in\includegraphics[width=0.49\textwidth]{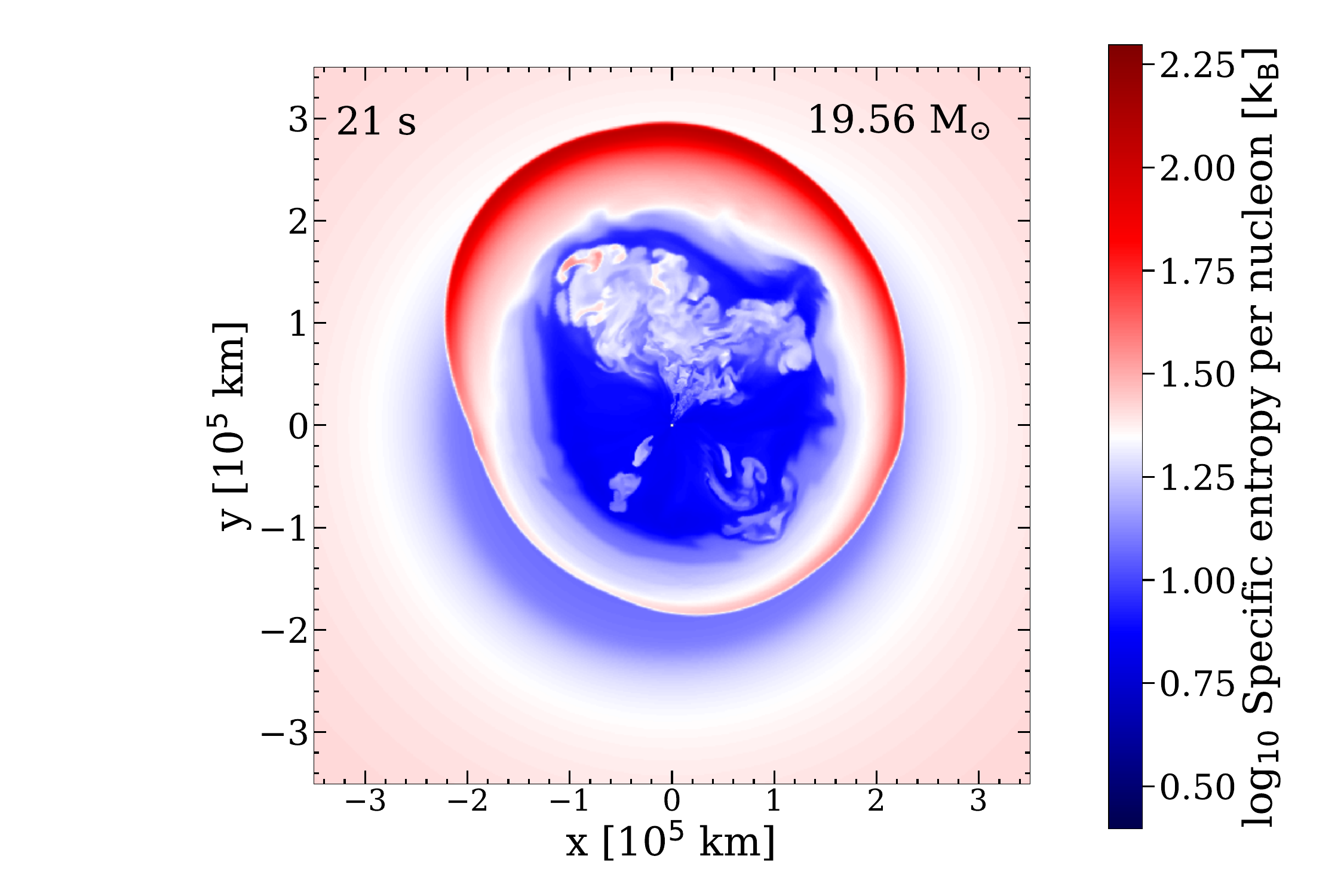}
    \hskip+0.0in\includegraphics[width=0.49\textwidth]{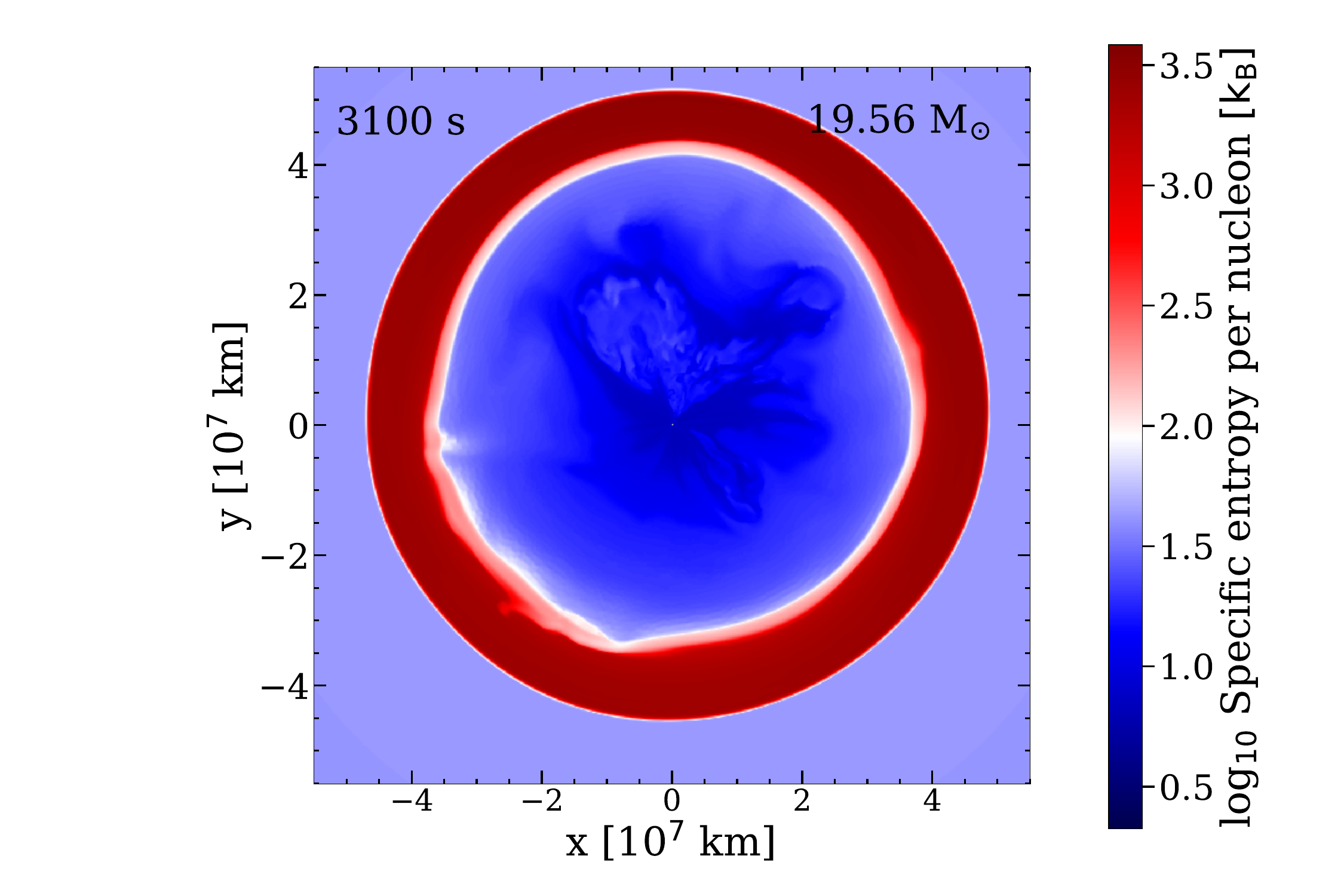}
    \hskip+0.0in\includegraphics[width=0.49\textwidth]{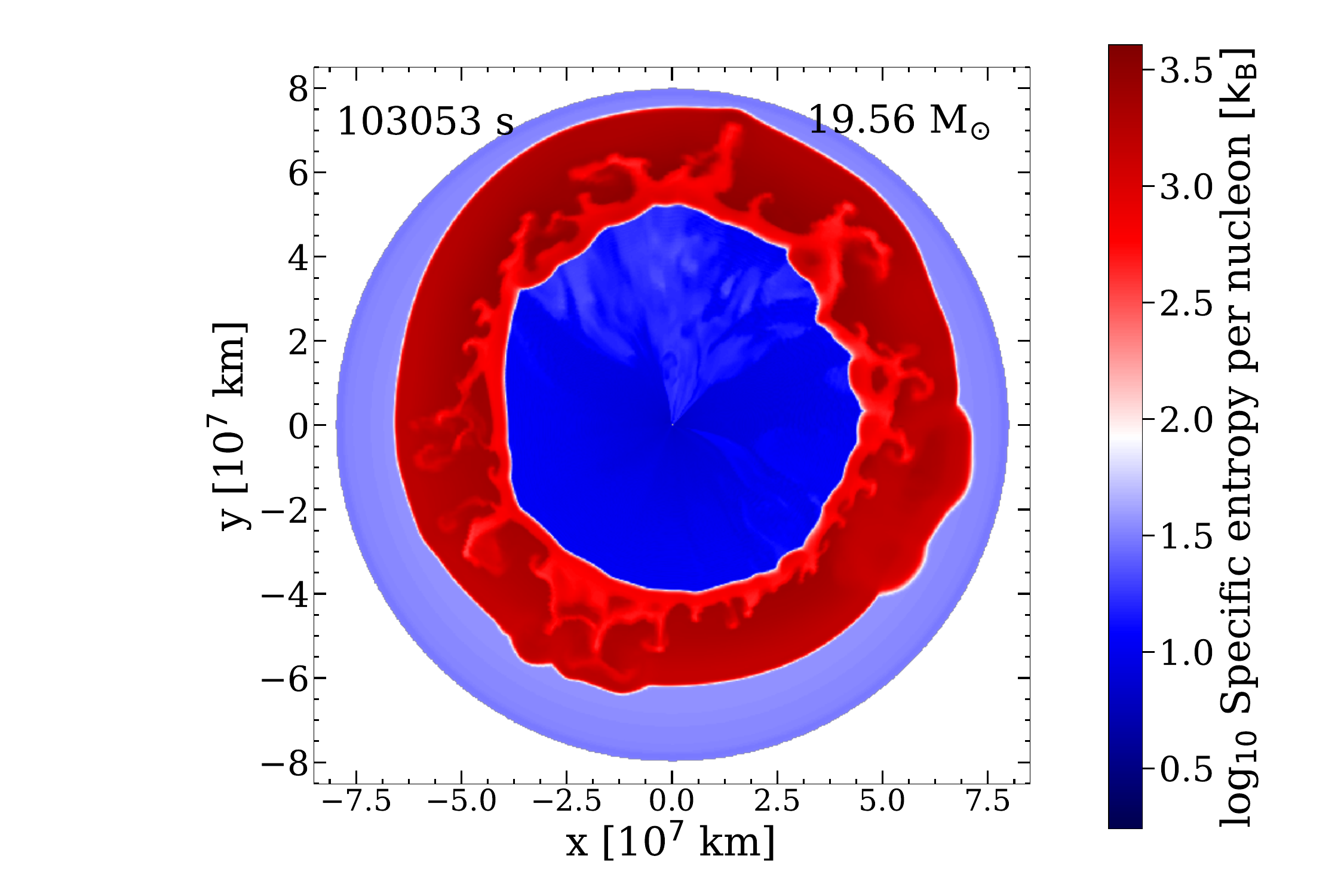}
    \caption{Entropy (per baryon per Boltzmann's constant) stills for the 19.56-M$_{\odot}$ model during the FLASH phase of the simulations. Within $\sim$1000 seconds after bounce, the majority of the fallback accretion has subsided and $\sim$one solar mass of material has accreted onto the compact object, leaving a black hole with a mass of $\sim$3.07 M$_{\odot}$. $\sim$10.56 $M_{\odot}$ is ejected in a vigorous explosion with an energy greater than 2.5 Bethe. Note the large changes in scale from still to still among the four panels, with the corresponding change in color maps and epoch after bounce (10, 21, 3100, and 103053 seconds). In the first 20 seconds, we see in the four plumes of ejecta a deformed shock front. The shock partially sphericizes as it expands into the envelope. Near shock breakout, we see extensive Rayleigh-Taylor fingers spanning the shock width.}
    \label{fig:before_BH_1956}
\end{figure} 

\subsection{{\bf Channel 1}: Black Hole Formation Accompanied by an Asymmetrical Vigorous Explosion: 40-Solar-Mass and 19.56-Solar-Mass/Solar-Metallicity Models}
\label{40_19.56_explosion}

Table \ref{table1} summarizes some of the properties and observables for the example simulations for all four black hole formation channels.  Included in this table are the corresponding numbers for many of our long-term 3D simulations that ostensibly left a neutron star and clearly exploded as a core-collapse supernova. Though the focus of this paper is on our examples for the four stellar-mass black hole formation channels we have identified, the broad comparison provided in Table 1 provides much needed context.

The two exemplars of this channel black hole formation that is accompanied by a high energy, asymmetric explosion are our 40- and 19.56-$M_{\odot}$ models. These models have the highest compactness of any of the solar-metallicity models we have yet simulated in 3D (see Table \ref{table1}). The left panel of Figure \ref{fig:rho_full} depicts the shallowness of their initial inner mass density profiles (the two just below the red 100-$M_{\odot}$ curve). Figure \ref{fig:rs} demonstrates that such high compactness models explode earlier after bounce than most of the others of the highish compactness class (``orange"), and are accompanied during the early explosive phase by a strong spiral SASI mode (\S\ref{basics}). Comparing their relative mass density profiles in the 2.0$-$3.0 $M_{\odot}$ interior mass interval with those just below them in density (and compactness), one can easily understand that the associated post-bounce mass accretion rates and accretion luminosities for the 40- and 19.56-$M_{\odot}$ models are higher. Figure \ref{fig:Qdot} demonstrates this clearly. The heating rate for them is eclipsed only by that for the 100-$M_{\odot}$ of Channel 3. Importantly, the models just below them in compactness and mass density in this interval explode energetically, but leave behind neutron stars (Table \ref{table1}). Those models benefit from high mass accretion rates and accretion luminosities, but possess a mass accretion rate that is low enough not to lead to a black hole, given the high associated explosion energies. Hence, Channel 1 lies in a special interval of mass profile and compactness that threads the needle between Channel 3, Channel 2, and neutron-star formation channels (see \S\ref{basics}).     

In \citet{burrows_40}, we described in detail the early characteristics of the 40-$M_{\odot}$ model out to 8.8 seconds after bounce.  We have now carried this model out to $\sim$6000 seconds and Table \ref{table1} has been updated accordingly. The right-hand panel of Figure \ref{fig:rho_full} portrays the evolution with ($\log_{10}$)time of the gravitational mass of our recent collection of 3D models, including all our examples of black hole formation (dashed curves). The model with the second highest residual mass at early times is the 40-$M_{\odot}$ model, the 19.56-$M_{\odot}$ is just below it, and the 100-$M_{\odot}$ is just above it. The filled circles indicate the time of black hole formation.
  
First, we notice the early hierarchies on this panel, including those models that leave neutron stars. As compactness goes up, so too does the residual mass. Most of the models form neutron stars (Table \ref{table1}). For those models that formed a black hole within seconds (40-, 19.56, and 100-$M_{\odot}$), the model that forms a black hole later does so at lower baryon and gravitational masses (cf. the 19.56-$M_{\odot}$ model). Notice for these models the decrease with increasing compactness in the time interval between explosion and black hole formation. The longer time reflects the relatively lower accretion rates, but also the longer time to cool and deleptonize. The result is an alteration in the entropy and $Y_e$ profiles, both of which have an effect on the critical baryon and gravitational masses for the general-relativistic instability to a black hole. We find that the corrections to the baryon mass at collapse to a black hole for the 19.56-, 40-, and 100-$M_{\odot}$ models are $\sim$0.1, $\sim$0.3, and $\sim$0.6 $M_{\odot}$, respectively. These are not small. The gravitational masses at these early stages are of course lower, but still follow the hierarchy at black hole formation as a function of compactness/$\dot{M}$.    

We refer to \citet{burrows_40} for a more detailed discussion of the early behavior of the 40-$M_{\odot}$ simulation. However, unanticipated in that paper was the significant fallback mass for this model after the F{\sc{ornax}} phase.  \citet{burrows_40} had estimated a final black hole mass of $\sim$3.5 $M_{\odot}$. Indeed, at $\sim$21 seconds after bounce, the mean shock radius was $\sim$200,000 km and the accretion rate into the interior appeared to be subsiding. However, the large binding energy of the outer mantle of helium and hydrogen for this model maintained fallback for the next few hours, with the result that asymptotically the gravitational mass of the black hole left behind was $\sim$9.0 $M_{\odot}$ (almost three times that seen in the \citet{burrows_40} paper at early times!). Figure \ref{fig:rho_full} depicts its evolution.  The late-time fallback mass accretion rate assumes the expected $\frac{1}{t^{5/3}}$ behavior. This resulted in an ejecta mass of $\sim$6.2 $M_{\odot}$. The explosion energy changed little and settled around $\sim$1.75 Bethes, but the fallback carried a fraction of the $\sim$0.17 $M_{\odot}$ of $^{56}$Ni produced into the interior, with the result that 0.114 $M_{\odot}$ of  $^{56}$Ni was eventually ejected.  Moreover, the recoil kick speed moderated to $\sim$550 km s$^{-1}$ \citep{burrows_kick_2024}. Hence, this explosion would look similar to a canonical supernova, but with a significant blast asymmetry.

We note that the kicks and spin-up in the fallback context are quite nuanced, given the aleatoric nature of fallback hydrodynamics \citep{janka2022,muller2023}. 
However, we still find that the kick speed of the 40-$M_{\odot}$ model's black hole is larger than the birth kicks inferred for the stellar-mass black holes in X-ray binaries \citep{2005ApJ...625..324W,2019MNRAS.489.3116A,oh.bh.kick} and almost guarantees that a fraction (of currently undetermined value) of Channel 1 black holes are unlikely to have companions. So, the residue of such an explosion would be a challenging observational target. 

Interestingly, if this 40-$M_{\odot}$ model had experienced binary mass transfer and been stripped to a ``stripped-envelope" structure, the fallback mass would have been minimal, the ejected $^{56}$Ni mass would have been more significant, the total ejected mass would have been much lower, as would the residual black hole mass, and the kick to the black hole would have been higher. This emphasizes the role of the reverse shock and mantle binding energy in the fallback process whose effects are contingent upon the presence and structure of the hydrogen envelope. In any case, the explosion energy would be little changed and would be similarly large. Hence, we \adam{speculate} that depending upon the degree of binary stripping (and likely the metallicity) this exemplar of black hole formation would result in a high explosion energy, but a spectrum of ejecta masses from a few to many $M_{\odot}$, an ejected $^{56}$Ni mass from high to lower, and a black hole masses up to perhaps $\sim$10.0 $M_{\odot}$. In all cases, a supernova explosion results. 

We generally get a large spin parameter (as high as $\sim$0.6)\footnote{\adam{The spin parameter ($a$) is defined as $\frac{L}{GM^2/c}$, where $L$ is the total angular momentum and $M$ is the core gravitational mass, and the other quantities have their usual meanings. The maximum possible $a$ is 1.0.}} that is very much a function of the character of the impact parameters of the (early)infall and (later)fallback plumes, themselves dependent on the chaotic flow field \citep{burrows_kick_2024}. The spin parameters of observed black-hole X-ray binaries \citep{2009ApJ...697..900M} are $\sim$0.01$-$0.05. Hence, we are not sure how robust this number is for such a model. This is just want we find. If true, such a high value would suggest that this model, if properly simulated with B-fields, could be a collapsar model for long-soft gamma-ray bursts \citep{1993ApJ...405..273W}. This is surprising since the initial model was non-rotating and the spin is induced \citep{rantsiou,burrows_kick_2024} during the early evolution.

Some of the final physical observables of the 19.56-$M_{\odot}$ model and simulation are given in Table \ref{table1}. This model has a much lower mantle binding energy than does the 40-$M_{\odot}$ model, with the result that fallback is much    less of a factor for it.  We see this in Figure \ref{fig:rho_full}. The model initially explodes asymmetrically with an energy of $\sim$2.5 Bethes, creating 0.228 $M_{\odot}$ of $^{56}$Ni. Black hole formation occurs at $\sim$3.8 seconds after bounce and, as a result, the neutrino heating phase of explosion lasts $\sim$3.5 seconds. The large heating rate depicted in Figure \ref{fig:Qdot} (the third highest), coupled with the long duration of neutrino driving, explains the high explosion energy. Figure \ref{fig:before_BH_19_1} portrays the approximately unipolar explosion morphology up to 3.767 seconds after bounce, that nevertheless dances in direction in response to fluctuating infalling low-entropy streams. Figure \ref{fig:before_BH_19} shows snapshots at various phases of the explosion from near onset to near black hole formation, but on a larger scale to show the early propagation of the shock wave. The differences between the flow character of the 40- and 19.56-$M_{\odot}$ explosions, though they are both energetic and asymmetrical, seem to have a random component due to the turbulent chaotic dynamics that characterizes the 3D CCSN mechanism. Clearly, more such high compactness models are necessary to fully determine the distribution functions of the various outcomes we are here witnessing only dimly with our specific ``19.56-$M_{\odot}$" and ``40-$M_{\odot}$" models. 

For the 19.56-$M_{\odot}$ model, the induced spin parameter ($a$) of the black hole at the time of its formation is only 0.006 \adam{(and at the end of the FLASH phase of the simulation is $\sim$0.09)}, the baryon mass (interior to 10$^{11}$ g cm$^{-3}$) is 2.278 $M_{\odot}$, and the corresponding gravitational mass is $\sim$2.071 $M_{\odot}$.  At black hole formation, we mapped the F{\sc{ornax}} model into FLASH and continued the calculation forward until $\sim$1.3$\times$10$^{5}$ seconds after bounce with a point mass and a diode inner boundary condition at 500 km. Stills of the subsequent evolution using FLASH are given in Figure \ref{fig:before_BH_1956}. Note that by the end of this series the shock had traversed half of the progenitor's radius and fallback has ceased. 

After the 19.56-$M_{\odot}$ model's explosion has asymptoted, the gravitational mass of the black hole residue is $\sim$3.12 $M_{\odot}$, the ejected $^{56}$Ni mass is $\sim$0.20 $M_{\odot}$, the explosion energy is still $\sim$2.5 Bethes, and the total ejecta mass is $\sim$10.5 $M_{\odot}$. The net jet asymmetry results in a large (asymptotic) recoil kick to the PNS of $\sim$1300 km s$^{-1}$, only $\sim$21 km s$^{-1}$ of which derives from the net neutrino impulse (calculated at the PNS surface). Approximately $\sim$0.03 $M_{\odot}$ of $^{56}$Ni had fallen back. The lower residual gravitational mass for the 19.56-$M_{\odot}$ model at black hole formation and terminally than witnessed for the 40-$M_{\odot}$ model at the corresponding phases reflects 1) the longer time to black hole formation (and the resulting higher explosion energy), itself a reflection of the slightly lower post-bounce $\dot{M}$s and (crudely) compactness for the 19.56-$M_{\odot}$ model; and 2) the larger mantle binding energy of the 40-$M_{\odot}$ model. Importantly, as noted earlier, due to the larger mantle binding energy for the 40-$M_{\odot}$ model, it experiences significantly more fallback than the 19.56-$M_{\odot}$ model. The final black hole mass of the 19.56-$M_{\odot}$ model \adam{is clearly in the putative lower black hole mass gap \citep{shao2022}} (similar to GW230529). Given its high kick speed, and depending upon the birth statistics, this subset of Channel 1 black holes would help make the observed lower mass gap in fact a true observational gap.
    
\adam{The lower final induced spin parameter for the 19.56-$M_{\odot}$ model ($\sim$0.09, as stated above)} seems related to the lower average impact parameters, which themselves may be slightly random. Again, an understanding of this difference awaits a more complete suite of 3D simulations to tease out the associated distributions that emerge from such chaotic behavior.

Therefore, we suggest that Channel 1 objects leave behind black holes, and result in respectable supernova explosion energies, significant $^{56}$Ni yields, a wide range of possible ejecta masses, and a spectrum of kicks from modest to large.  Recall that these specific quantitative outcomes depend upon the nuclear equation of state employed. We surmise that a nuclear equation of state with a higher maximum cold neutron star mass would result in a more energetic explosion and a lower final black hole mass. However, this too remains to be demonstrated.



\begin{figure}
    \centering
    \includegraphics[width=0.45\textwidth]{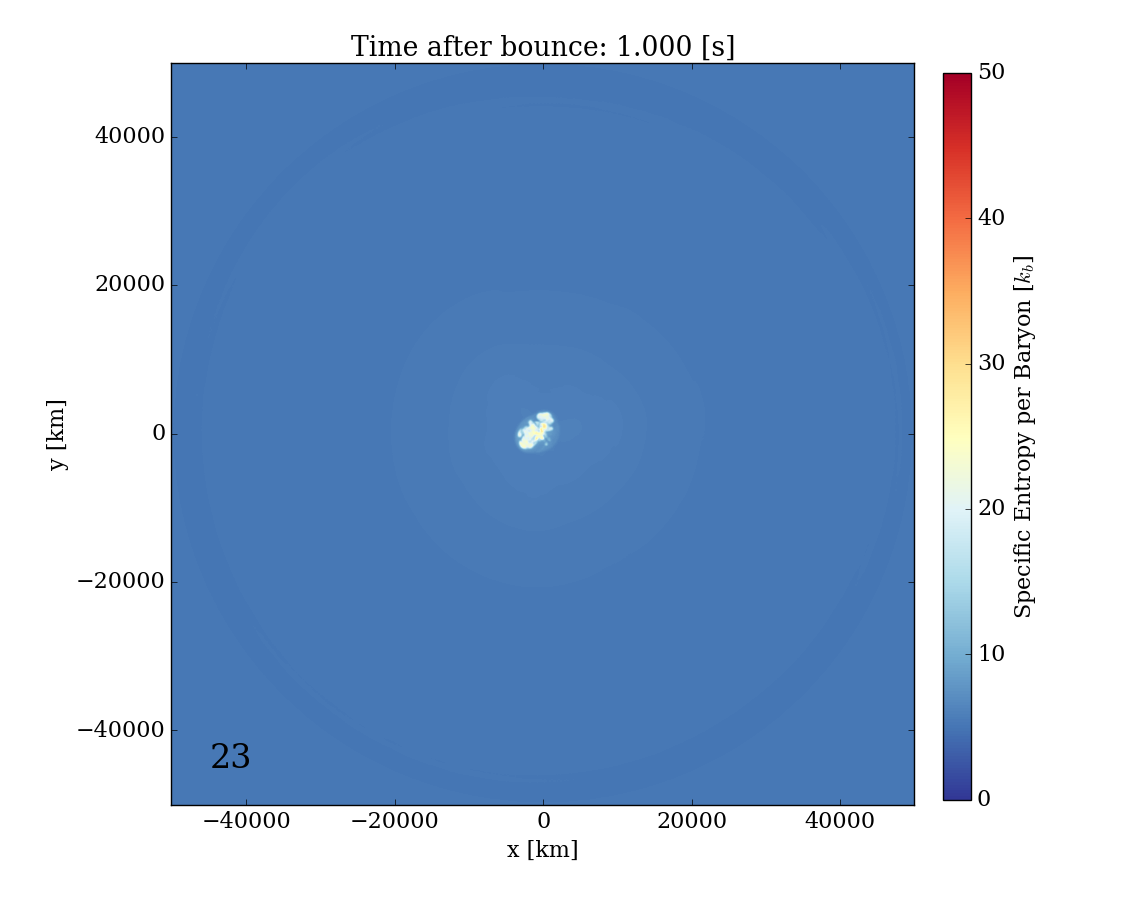}
    \includegraphics[width=0.45\textwidth]{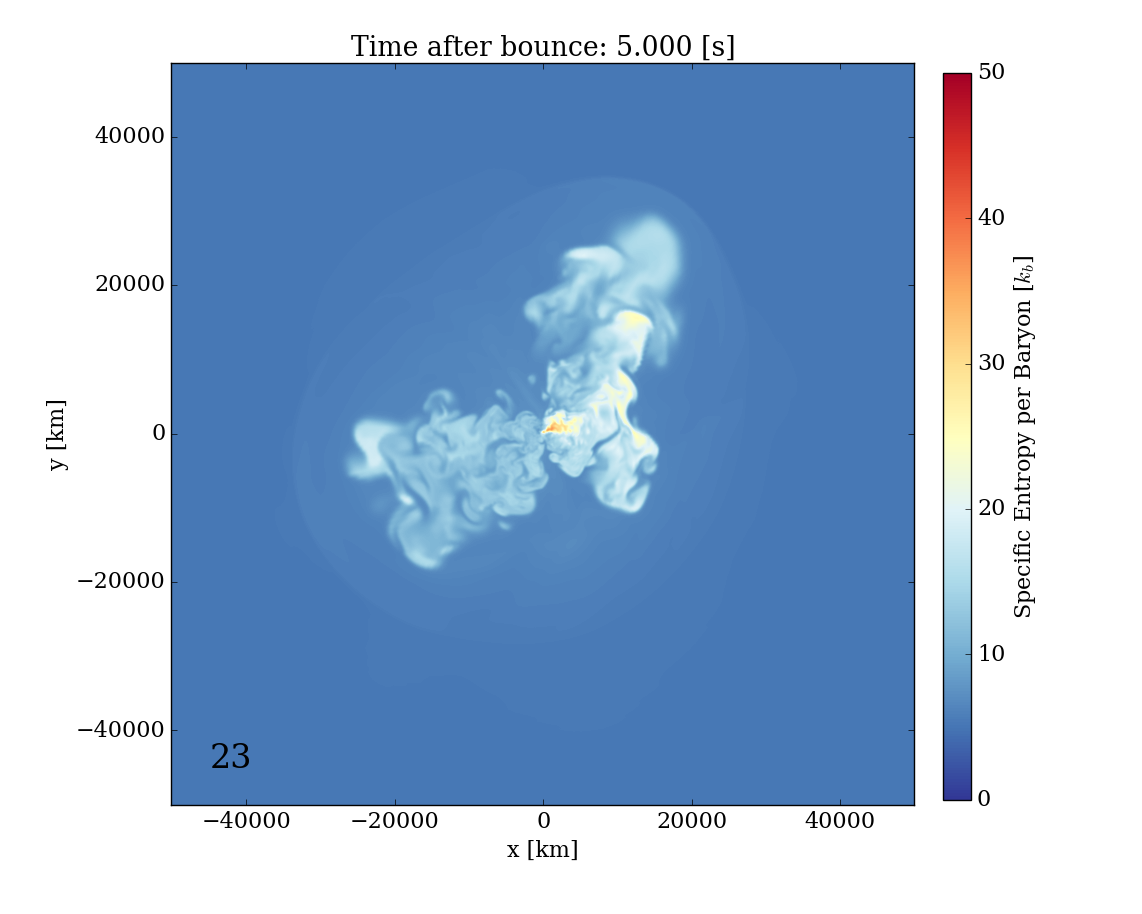}
    \hskip-0.0in\includegraphics[width=0.49\textwidth]{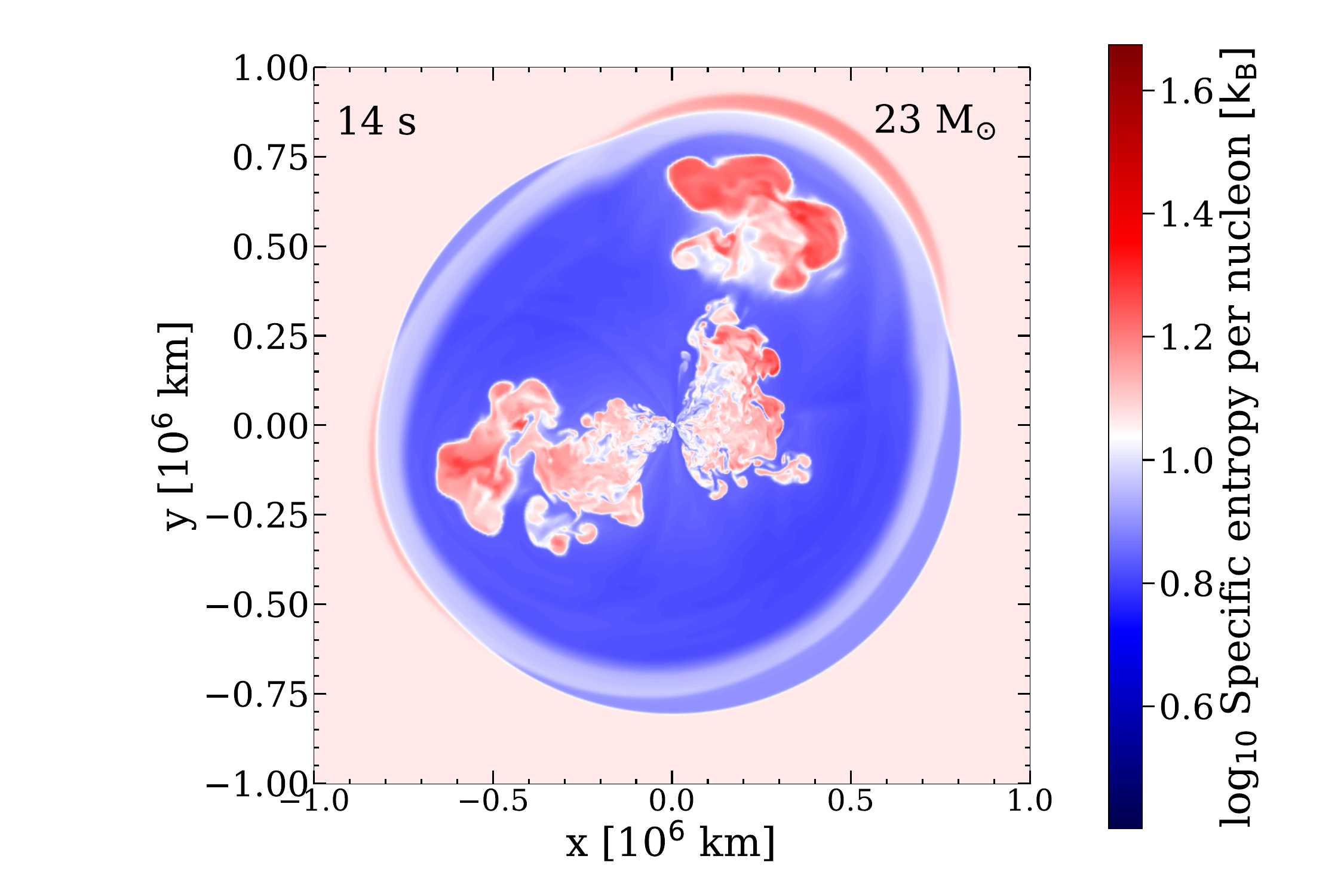}
    \hskip-0.0in\includegraphics[width=0.49\textwidth]{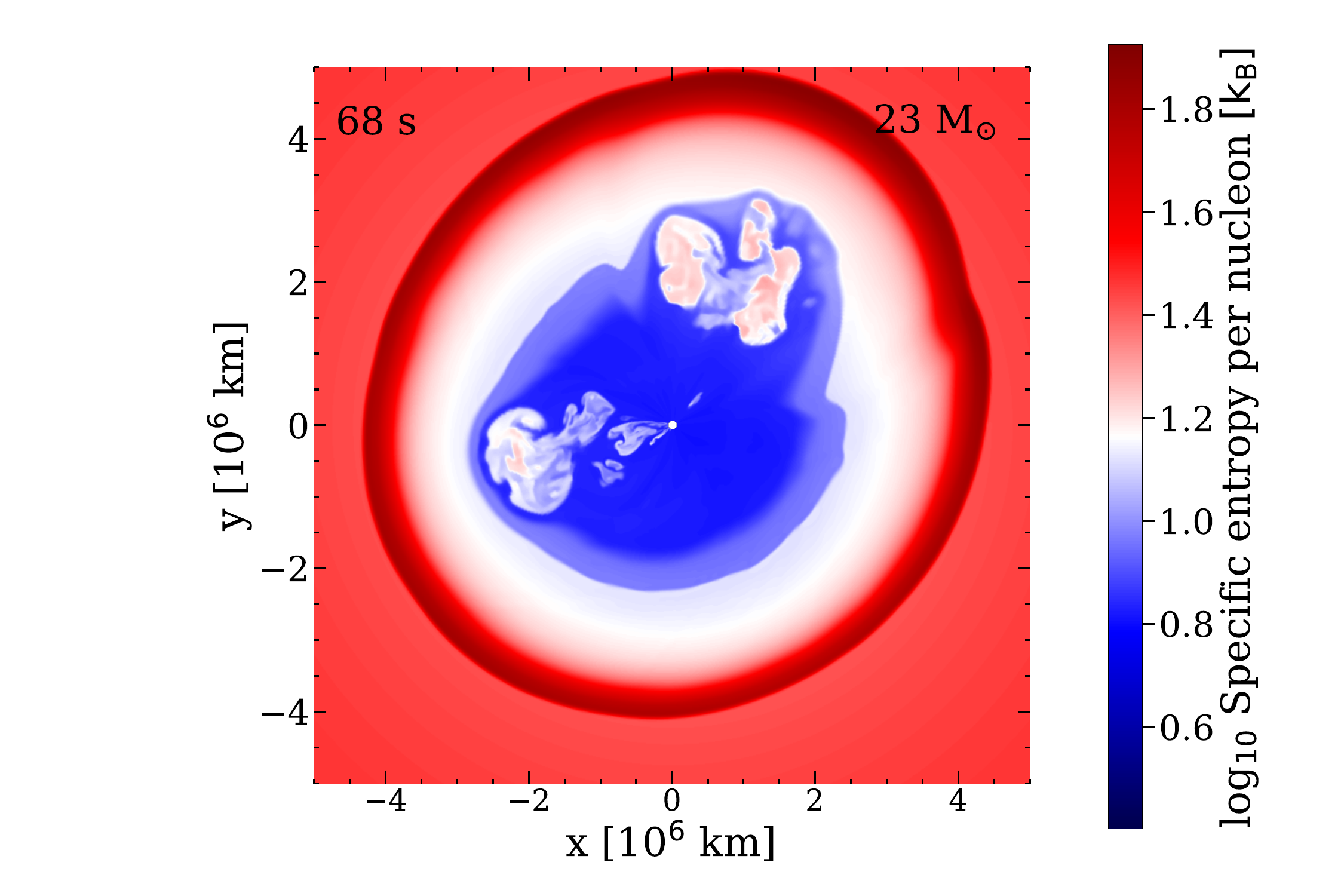}
    \hskip-0.0in\includegraphics[width=0.49\textwidth]{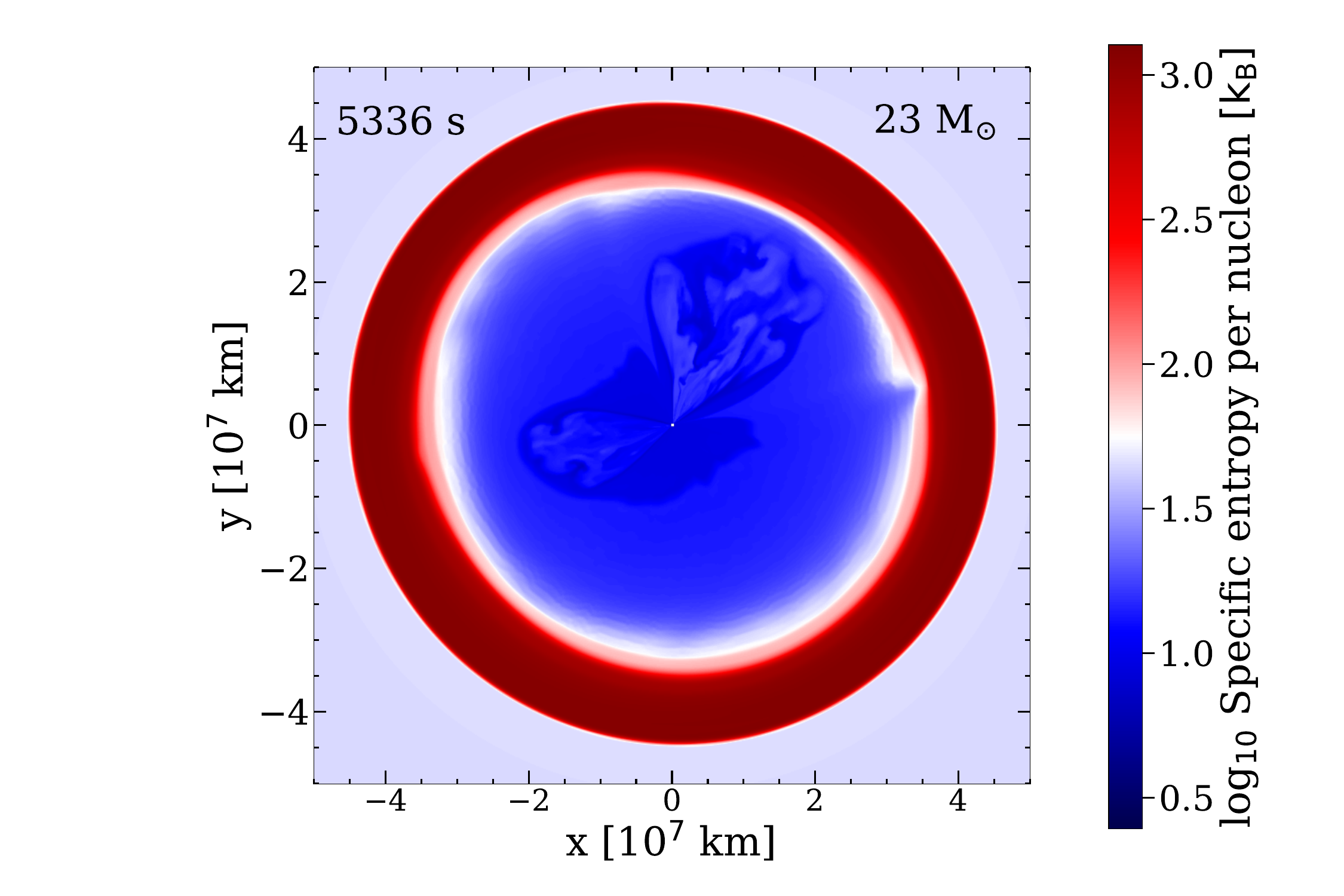}
    \hskip+0.0in\includegraphics[width=0.49\textwidth]{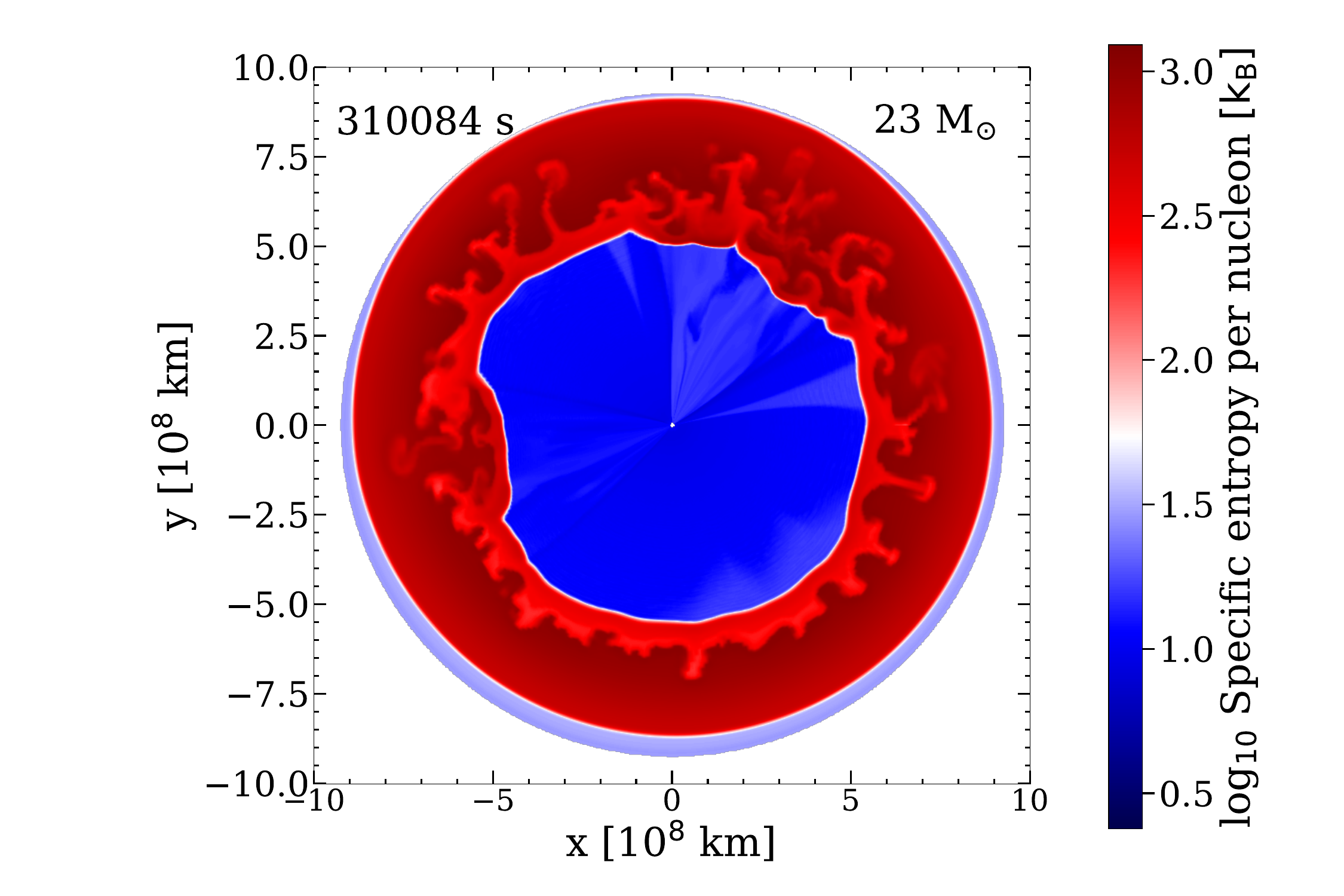}
    \caption{Sample snapshots of x-y slices in entropy space (per baryon per Boltzmann's constant) of the later explosion of the 23-$M_{\odot}$ solar-metallicity model for six different times after the model was mapped from F{\sc{ornax}} (top two) to FLASH (bottom four). Within $\sim$1000 seconds after bounce, the mass cut between the ejecta and the fallback matter destined to be incorporated into the black hole was determined.  A total of $\sim$4.9 $M_{\odot}$ is incorporated into the black hole, while $\sim$9.9 $M_{\odot}$ is ejected in a supernova explosion with an energy of $\sim$0.45 Bethes. Note the large changes in scale from still to still among the last four panels, with the corresponding change in color maps and time after bounce (1, 7, 14, 68, 5336, and 310084 seconds). The last still is after the shock breakout from the progenitor star and the onset of Rayleigh-Taylor-like instabilities at the hydrogen/helium interface. See text for a discussion.}
    \label{fig:before_BH_23}
\end{figure}

\begin{figure}
    \centering
    \includegraphics[width=0.49\textwidth]{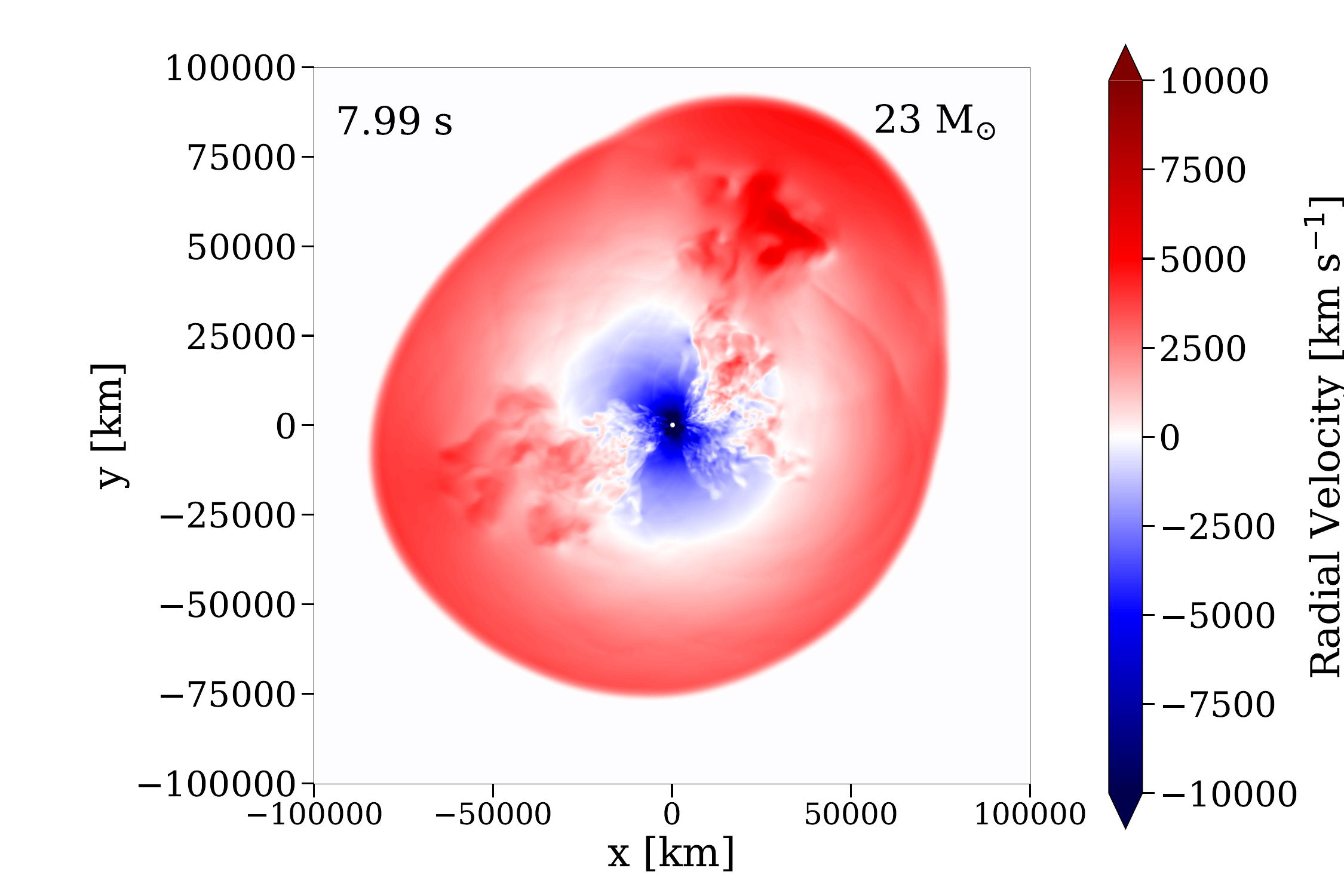}
    \includegraphics[width=0.49\textwidth]{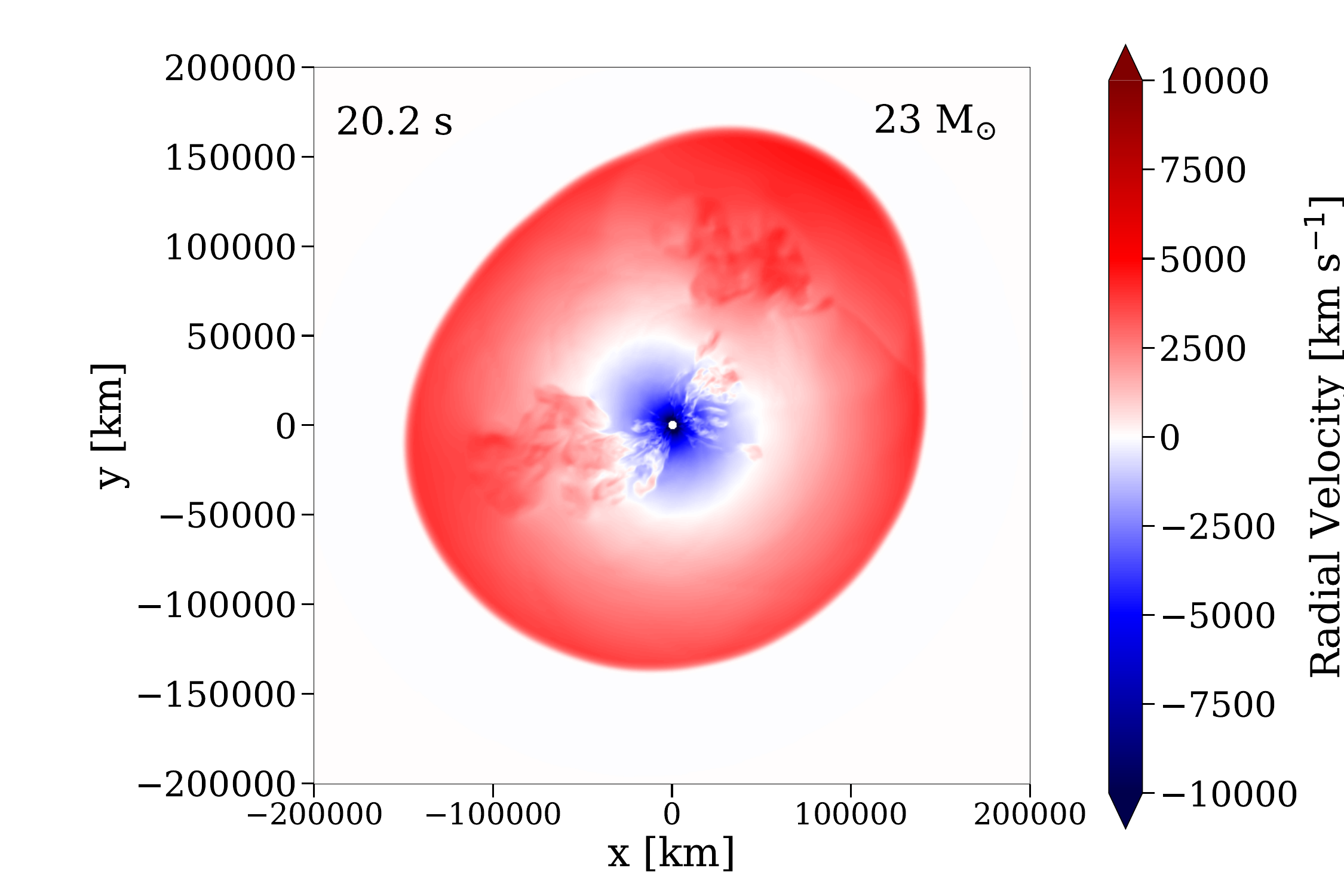}
    \caption{Two x-y slices of the velocity field during the explosion of the 23-$M_{\odot}$ solar-metallicity model at 7.99 and 20.2 seconds after bounce. The red field depicts the region were the matter at these times have positive radial velocity, while the blue field has negative radial velocity and the white fields are stagnating. Note the change of scale. This phenomenon can't be captured in 1D and reflects the presence for this model of significant early infall and later fallback.}
    \label{fig:before_BH_23_vel}
\end{figure}

\subsection{{\bf Channel 2:} Black Hole Formation in the Context of Fallback and a Supernova Explosion}
\label{bh_gap_weak}

In \citet{burrows_2020} and \citet{burrows_correlations_2024} we had concluded after 6.228 seconds of post-bounce simulation that our 23-$M_{\odot}$ model had exploded and left behind a neutron star with a gravitational mass of 1.722 $M_{\odot}$ (a baryon mass of 1.959 $M_{\odot}$). However, in retrospect we note that this model had a local minimum in $\xi_{1.75}$ among those in the \citet{sukhbold2018} cohort with which it clustered in progenitor mass and had lower mass density exterior to its silicon/oxygen interface (see Figure \ref{fig:rho_full}). Its post-explosion accretion $\dot{M}$ and heating rate (Figure \ref{fig:Qdot}) were correspondingly lower, with the result that the explosion energy we found then was half of that of the other members of this cohort \citep{burrows_correlations_2024}. With this in mind, we mapped our F{\sc{ornax}} run into FLASH, using the whole star and continued the calculation. Figure \ref{fig:before_BH_23} depicts in six panels (two during the F{\sc{ornax}} run and four during the FLASH run) the subsequent evolution out to 3.1$\times$10$^{5}$ seconds. The blast indeed continues, the shock wave breaks out of the star, and a supernova results.  However, within $\sim$50 seconds of bounce another $\sim$0.45 $M_{\odot}$ had joined the residual core and after $\sim$1000 seconds that augmentation had grown to $\sim$2.4 $M_{\odot}$, leaving after $\sim$300,000 seconds a $\sim$4.9 $M_{\odot}$ black hole. This is similar to the low masses inferred for the black holes in the SS433 \citep{picchi2020_ss433} and GRO J0422 \citep{gelino.2003_0422} X-ray binary systems. Figure \ref{fig:before_BH_23_vel} depicts snapshots of the simultaneous accretion and explosion of this model early during its evolution. The final explosion energy of the ejected $\sim$9.9 $M_{\odot}$ of the $\sim$15 $M_{\odot}$ remaining of the 23-$M_{\odot}$ ZAMS progenitor was $\sim$0.46 Bethes. 

Interestingly, of the 0.088 $M_{\odot}$ of $^{56}$Ni that was explosively generated in this 23-$M_{\odot}$ model \citep{wang_nucleo_2024,burrows_correlations_2024}, $\sim$0.039 $M_{\odot}$ was eventually ejected. This was due to the outer fraction of the $^{56}$Ni pockets that acted like bullets and decelerated less than the inner fraction, thereby penetrating into the stellar mantle. Moreover, though we had originally determined that this model had a kick of $\sim$280 km s$^{-1}$ \citep{burrows_kick_2024}, with fallback the net kick is partially due to the neutrino component, which then translates for this model into a black hole kick of $\sim$90 km s$^{-1}$.  However, there is no reason in principle that the net kick couldn't be larger or smaller. \adam{The corresponding final spin parameter ($a$) for this model is $\sim$0.135.} 

Therefore, what we find is a model that creates a $\sim$4.9 $M_{\odot}$ black hole, but explodes like a regular supernova with a modest \adam{(by which we mean in the mid-range of inferred measured values and our theoretical results)} explosion energy and $^{56}$Ni yield and a low recoil kick speed. This is not a ``silent" supernova! For slightly greater explosion energies and/or less bound envelopes, the residual black hole could have been lighter and the kicks could have been higher. This variation might arise from a different CCSN explosion code and different massive-star progenitor structures.  The converse is also true, wherein a larger mass black hole, accompanied by a still weaker explosion, might result. Given this, we can speculate that this channel of black hole formation might        yield black holes with \adam{a range of masses, yet to be determined.} Importantly, a set of similar exploding fallback scenarios was explored by \citet{chan2020}.

As we noted in \S\ref{40_19.56_explosion}, if this model had experienced binary stripping, the residual black hole mass, ejected mass, and $^{56}$Ni yield could have been different. Indeed, with significant stripping this same model might have given birth to a neutron star instead of a black hole, emphasizing the potential importance of binary interaction in determining the outcome of collapse. We emphasize that our numbers arise from the marriage of our supernova code with {\it KEPLER} progenitors at solar-metallicity. Changing the 3D CCSN code, stellar evolution module, mass loss prescription (for a singlet or a binary), and metallicity can change the numbers. However, we suggest that the existence of this channel of black hole formation with \adam{intermediate} energy and low kicks may well be robust. It is merely the mapping of progenitor ZAMS mass and metallicity to black hole mass, explosion energy, kick speed, and $^{56}$Ni yield that needs honing. 

What fraction of the more massive progenitors above $\sim$20 $M_{\odot}$ might follow this black hole channel remains to be determined, but it might not be small. Moreover, the existence of this channel, along with Channel 1, strongly suggests that some observed supernovae may well have birthed black holes. In addition, though we find small kicks for this channel, if the black hole progenitor is in a wide zero-eccentricity binary, the Blaauw mechanism \citep{Blaauw,Blaauw2} could leave behind an eccentric binary if the explosion is slightly weaker and/or the residual progenitor mass is slightly lower than we find here for our 23-$M_{\odot}$ model (see Table \ref{table1}). Hence, binary black holes such as Gaia BH1 and BH2 might be associated with Channel 2. Clearly, it will be important to determine if there are observational signatures in the supernova measurements that might discriminate this channel and this should be the subject of future research.


   
\begin{figure}
    \centering
    \hskip-0.0in\includegraphics[width=0.49\textwidth]{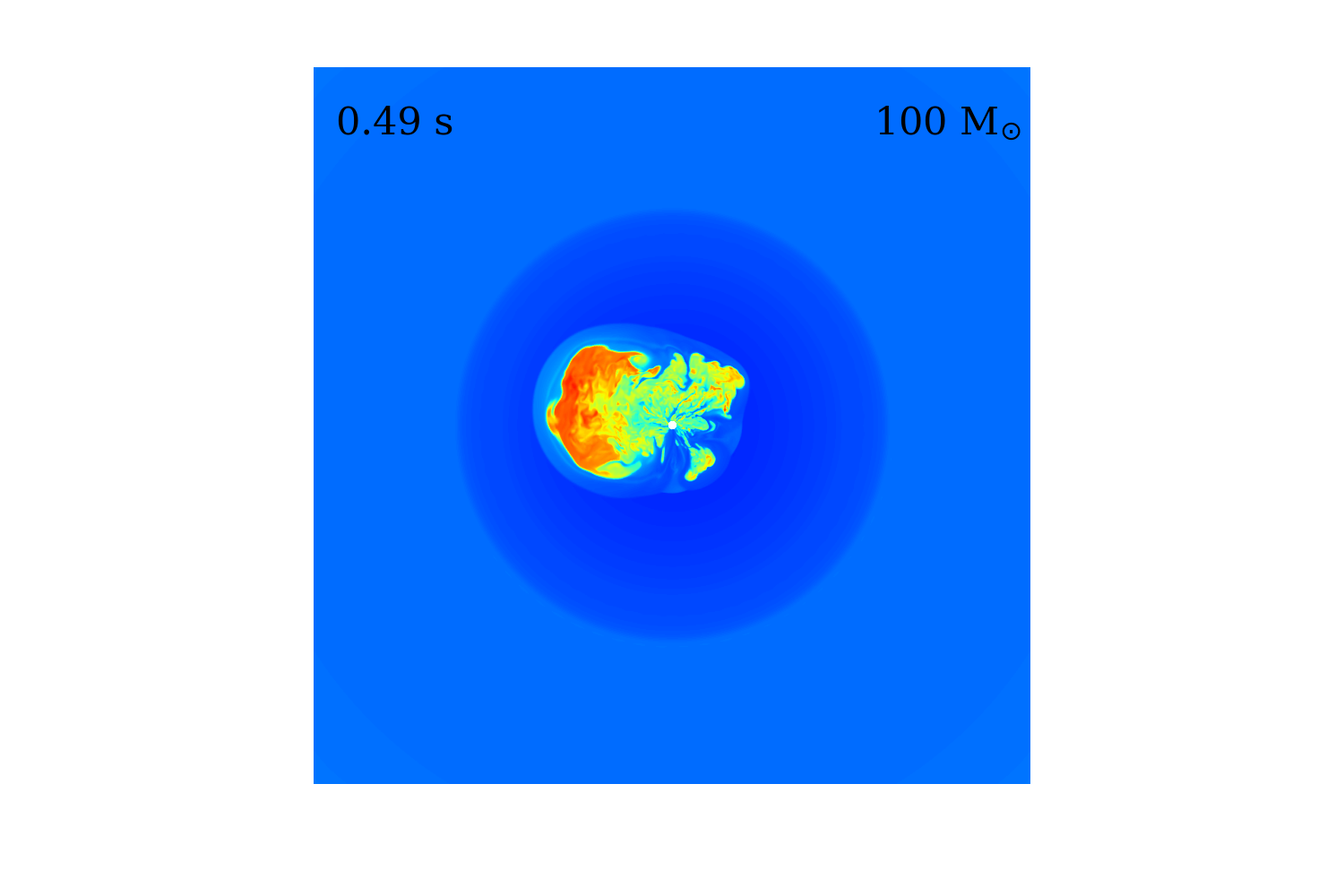}
    \hskip-0.0in\includegraphics[width=0.49\textwidth]{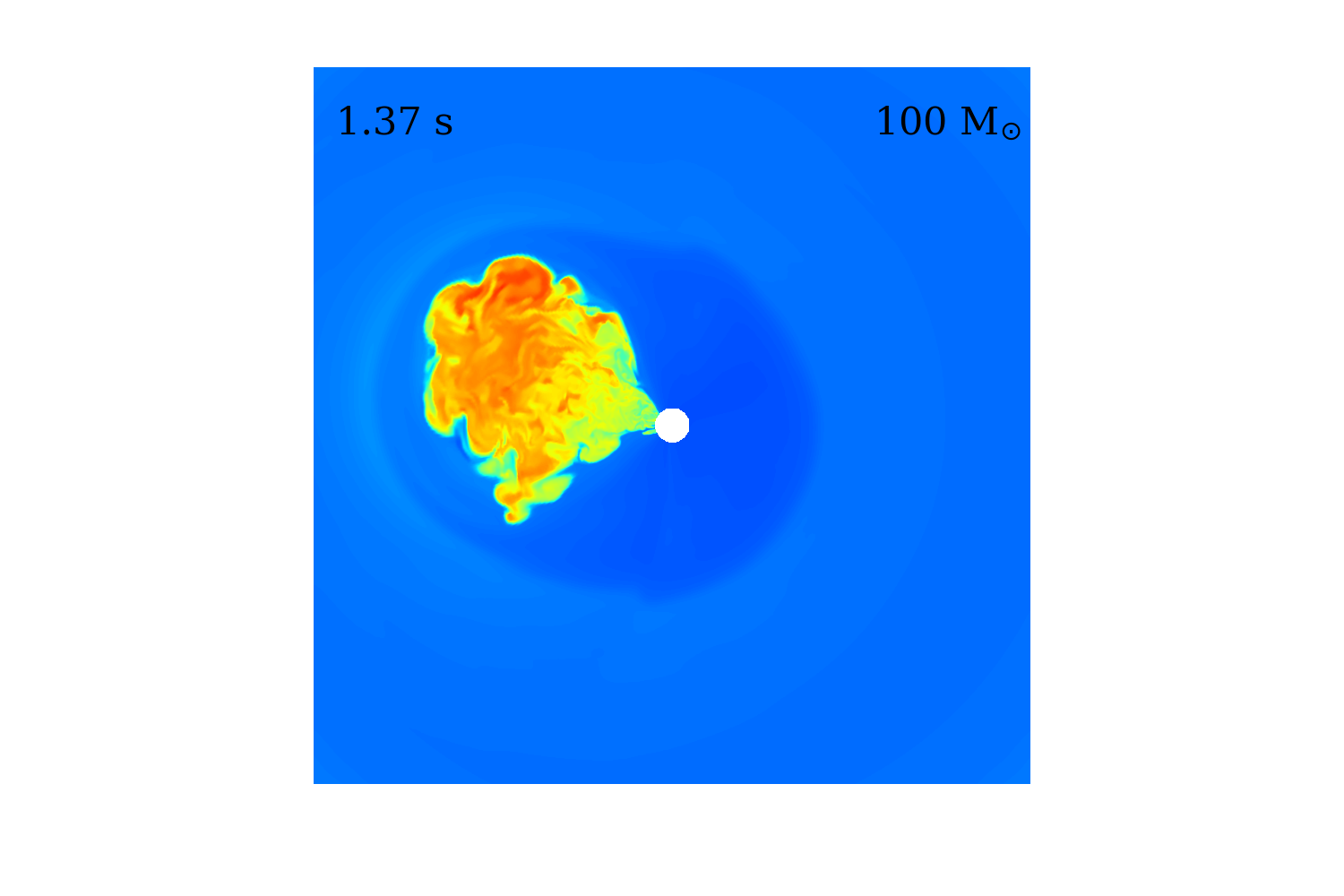}
    \hskip-0.0in\includegraphics[width=0.49\textwidth]{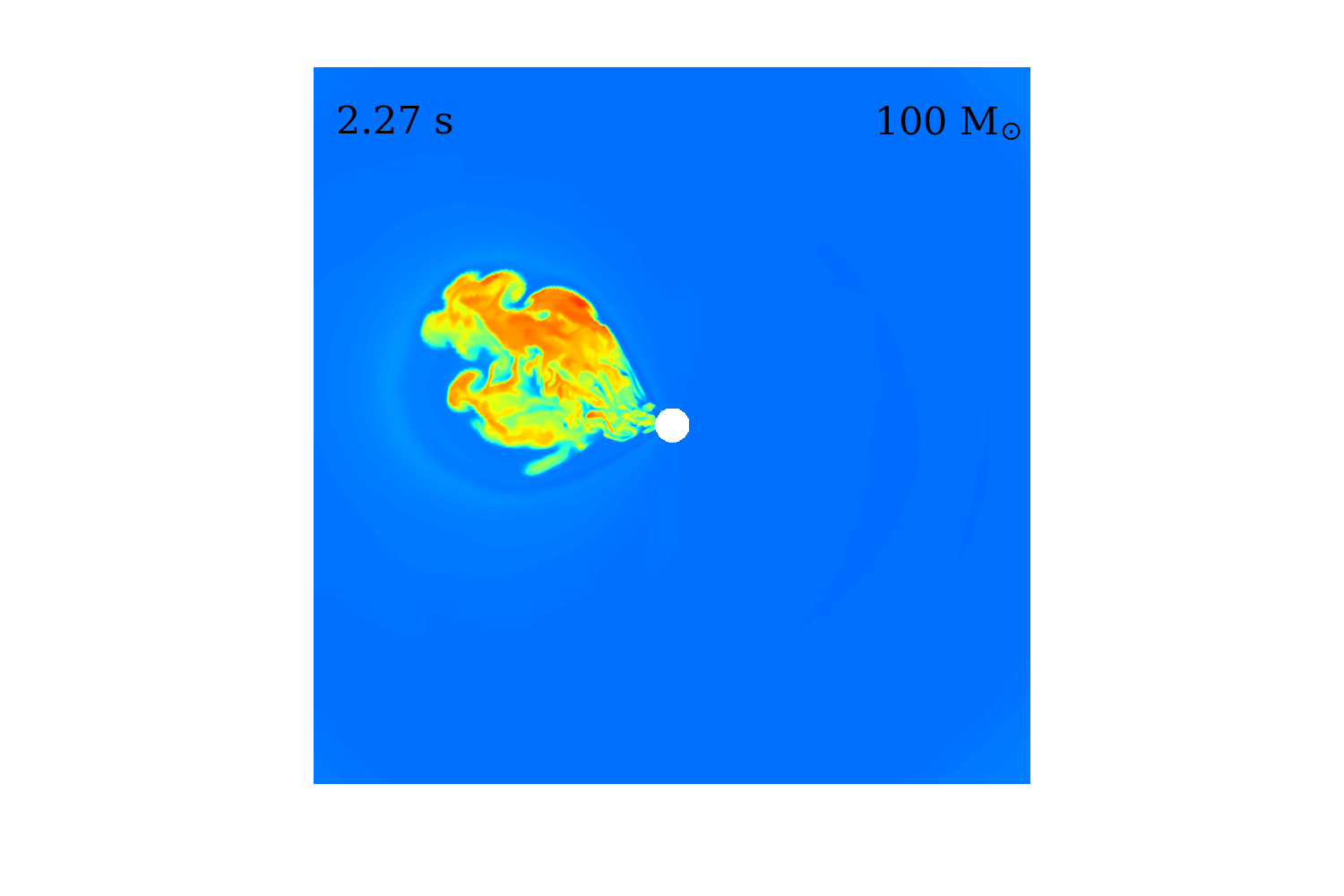}
    \hskip+0.0in\includegraphics[width=0.49\textwidth]{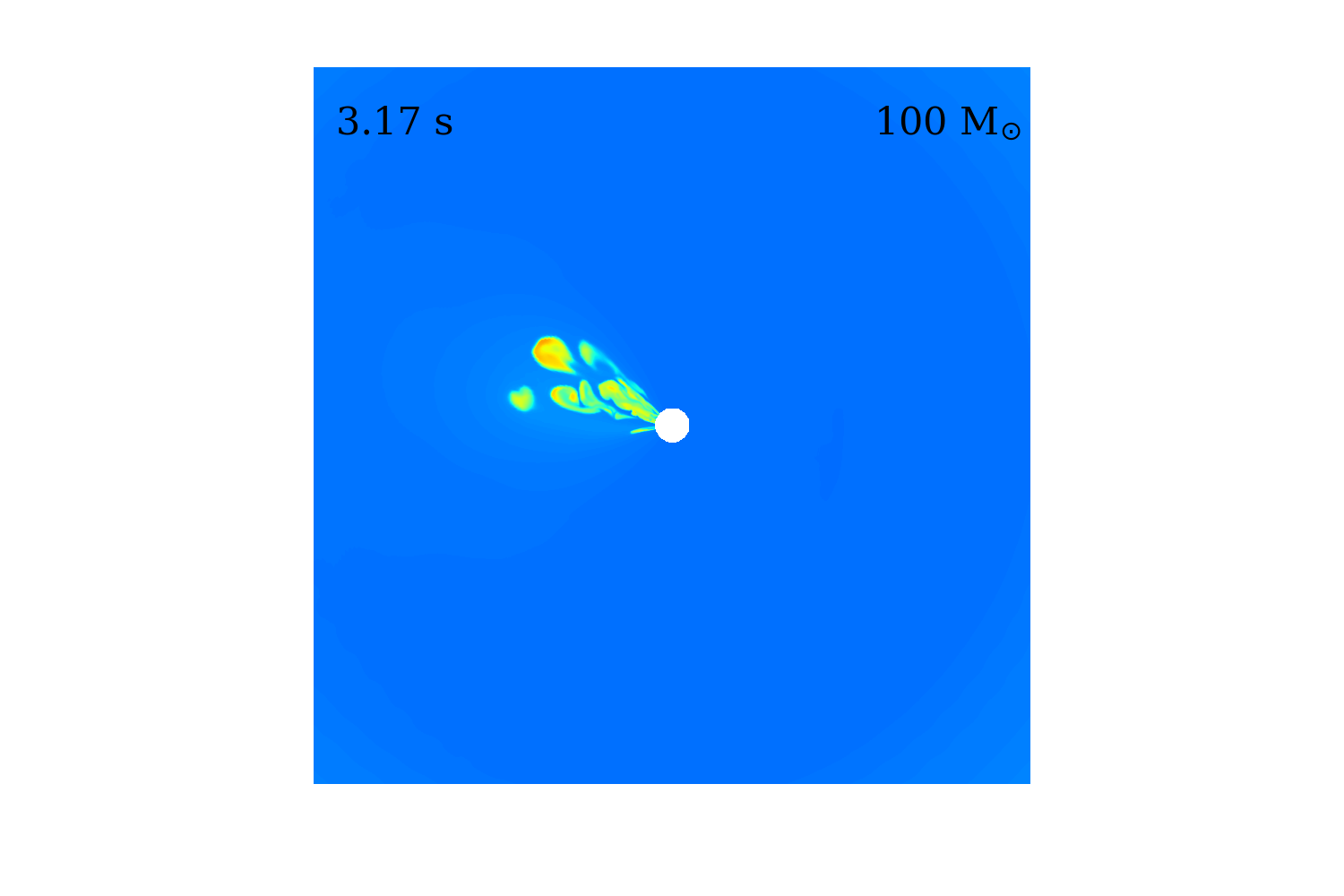}
    \caption{Sample snapshots of x-y slices in entropy space (per baryon per Boltzmann's constant) of the early explosion of the tenth-solar-metallicity 100-$M_{\odot}$ for four different times after bounce. Red is higher entropy and blue is the lower entropy of the progenitor matter into which the blast initially runs. The horizontal scale is 22,000 km. The upper left panel depicts the doomed explosion just $\sim$50 milliseconds after the black hole forms, which happens at $\sim$0.44 seconds after bounce. The dark-blue/light-blue transition in the top left and top right panels is at the oxygen/carbon interface. At the time of the top right panel the shock wave has penetrated this interface and for the bottom two panels it has left the graphical domain. After $\sim$10 seconds (not shown) the shock has weakened into a sound wave. As the bottom two panels suggest, within mere seconds the ejecta start to stall and fallback. Within $\sim$50 seconds, a $\sim$37 $M_{\odot}$ black hole has been created in the interior. See text for a discussion.}
    \label{fig:before_BH_100}
\end{figure}

\subsection{{\bf Channel 3:} A Pulsational-Pair-Instability Progenitor: 100 Solar Masses}
\label{100_tenth}

To capture the basic hydrodynamic evolution and outcome of this
channel of black hole formation, we mapped a model of the core of a 100 $M_{\odot}$ star at one-tenth solar metallicity at its terminal stages (kindly provided by Stan Woosley, private communication, generated using the {\it KEPLER} code) into our radiation/hydrodynamics code F{\sc{ornax}} \citep{skinner2019}. {\it KEPLER} witnessed three dynamical pulsations of the stellar envelope just prior to the onset of core collapse in a pulsational-pair instability supernova with an ejecta kinetic energy of 
$\sim$$4.3\times 10^{50}$ ergs (S. E. Woosley, in preparation). Our goal was to determine whether a vigorous core-collapse supernova might accompany the PPISN and contribute to its light curve, if only secondarily, and to determine the mass of the black hole produced. A tertiary goal was to explore the behavior and signatures of this channel of black hole formation.

The collapse, bounce, and subsequent evolution initially echoed the behavior seen for the 40$M_{\odot}$ ZAMS progenitor (see \S\ref{40_19.56_explosion} and \citet{burrows_40}). As Figure \ref{fig:rs} demonstrates, the shock stalls, its mean radius peaks within $\sim$70 milliseconds near $\sim$180 kilometers, and it then settles lower.  However, as the right-hand panel of Figure \ref{fig:rs} shows, a spiral SASI mode quickly emerges and the degree of asphericity of the stalled shock grows. The shorter time to this instability reflects the much larger compactness and $\dot{M}$ of this progenitor (see Table \ref{table1} and the left-hand panel of Figure \ref{fig:rho_full}). The combination of the diminution of the accretion ram pressure along a wobbling direction, the enlargement of the gain region, and the high accretion luminosity for such a high compactness core kicks the core into a very asymmetrical explosion at $\sim$250 milliseconds after bounce. At this time, the baryon mass of the PNS core is $\sim$2.7 $M_{\odot}$ (see right panel of Figure \ref{fig:rho_full}) and the core continues to accrete mass (mostly along the antipodal direction to the bulk of the explosion) at a rate of $\sim$2.0 $M_{\odot}$s$^{-1}$.

These near-antipodal accretion streams are post-explosion ``infall" (see \S\ref{basics}), since though they encountered the shock wave their speeds did not change sign in the process. At $\sim$400 milliseconds after bounce, along the most vigorously exploding direction the shock speed has reached $\sim$22,000 km s$^{-1}$, while the slowest shock speed (near the opposite side) is only $\sim$8000 km s$^{-1}$. However, at $\sim$440 milliseconds after bounce, the central object has reached a baryon mass of $\sim$2.72 $M_{\odot}$ and within 1 millisecond the central general-relativistic metric lapse reaches $\sim$0.42 and the core forms a black hole.  We see this event indicated in the right panel of Figure \ref{fig:rho_full}. The F{\sc{ornax}} code crashes at this point, but we then map the flow exterior to 100 kilometers onto the hydrodynamic code FLASH \citep{FLASH2000,FLASH2012,vartanyan_breakout_2024}, place a point mass with the PNS's current gravitational mass in the center, and continue the simulation using FLASH. 

The interval between explosion ($\sim$250 milliseconds) and black hole formation ($\sim$440 milliseconds) for the 100 $M_{\odot}$ model was less than $\sim$200 milliseconds.  This contrasts with the $\sim$1.5-second duration for the 40-$M_{\odot}$ model \citep{burrows_40} and the $\sim$3.5 duration for the 19.56 $M_{\odot}$ model (\S\ref{40_19.56_explosion}).  In both the latter cases, there was sufficient time for the high neutrino absorption rate associated with the high post-explosion infall accretion luminosity (due to the high compactness in their stellar mantles and the breaking of spherical symmetry) to unbind some or most of the residual mass of the progenitor star. The result in those cases, discussed in \S\ref{40_19.56_explosion}, was an energetic asymmeterical supernova explosion \adam{with a black hole residue} and a high recoil kick.

However, for the 100-$M_{\odot}$/one-tenth-solar model, the even higher compactness of its envelope (Table \ref{table1}) and greater $\dot{M}$s, while they led to shock revival and a vigorous start due to the associated high accretion luminosity and neutrino heating rate, also led to the more rapid accumulation of gravitational mass in its core and an early collapse into a black hole. The associated truncation of neutrino heating quickly aborted the driving of the explosion and left the blast with insufficient energy to unbind the envelope.

Using FLASH, we followed the shock wave through the oxygen core. Within four seconds of bounce most of the originally infalling plumes had been accreted. After $\sim$50 seconds, the shock wave had degenerated into a weak sound wave and merged into the flow.  This is similar to what \citet{rahman2022} witnessed for their models. By this time, accretion onto the newly-formed black hole had all but ceased, leaving a black hole at the center with a mass of $\sim$37 $M_{\odot}$.  This is most of the mass on the grid and implies that, ignoring secondary explosive phenomena (see \S\ref{silent}) due to the ingestion of a whole star into a geometrically small black hole, there is no additional contribution to the PPISN due to such associated core collapse. Figure \ref{fig:before_BH_100} portrays the early explosion, fizzling, and fallback during the first post-bounce $\sim$3 seconds of the core evolution of this model. Note the extreme asphericity of the initial blast, and its dramatic reversal. 

Finally, the total recoil kick just before black hole formation is $\sim$423 km s$^{-1}$, $\sim$14 km s$^{-1}$ of which is due to neutrinos.  Within 100 seconds, after the $\sim$37 $M_{\odot}$ black hole has been assembled and assuming there was no anisotropic mass ejection due to the accumulation of so much mass from such large radii, that recoil is nullified. The final recoil would then be $\sim$1 km s$^{-1}$. The difference of this number from zero is certainly lost in the inevitable errors of our overall simulation effort. However, as \citet{antoni} have suggested there is likely to be a mild secondary explosive effect, with perhaps some degree of asymmetry, due to the infall of a turbulent zone from such large radii. However, this feature has yet to be properly quantified. Moreover, the gravitational memory of this aborted CCSN explosion should survive in the low-frequency component of its gravitational-wave signature \citep{choi2024}.

\begin{figure}
    \centering
    \includegraphics[width=0.80\textwidth]{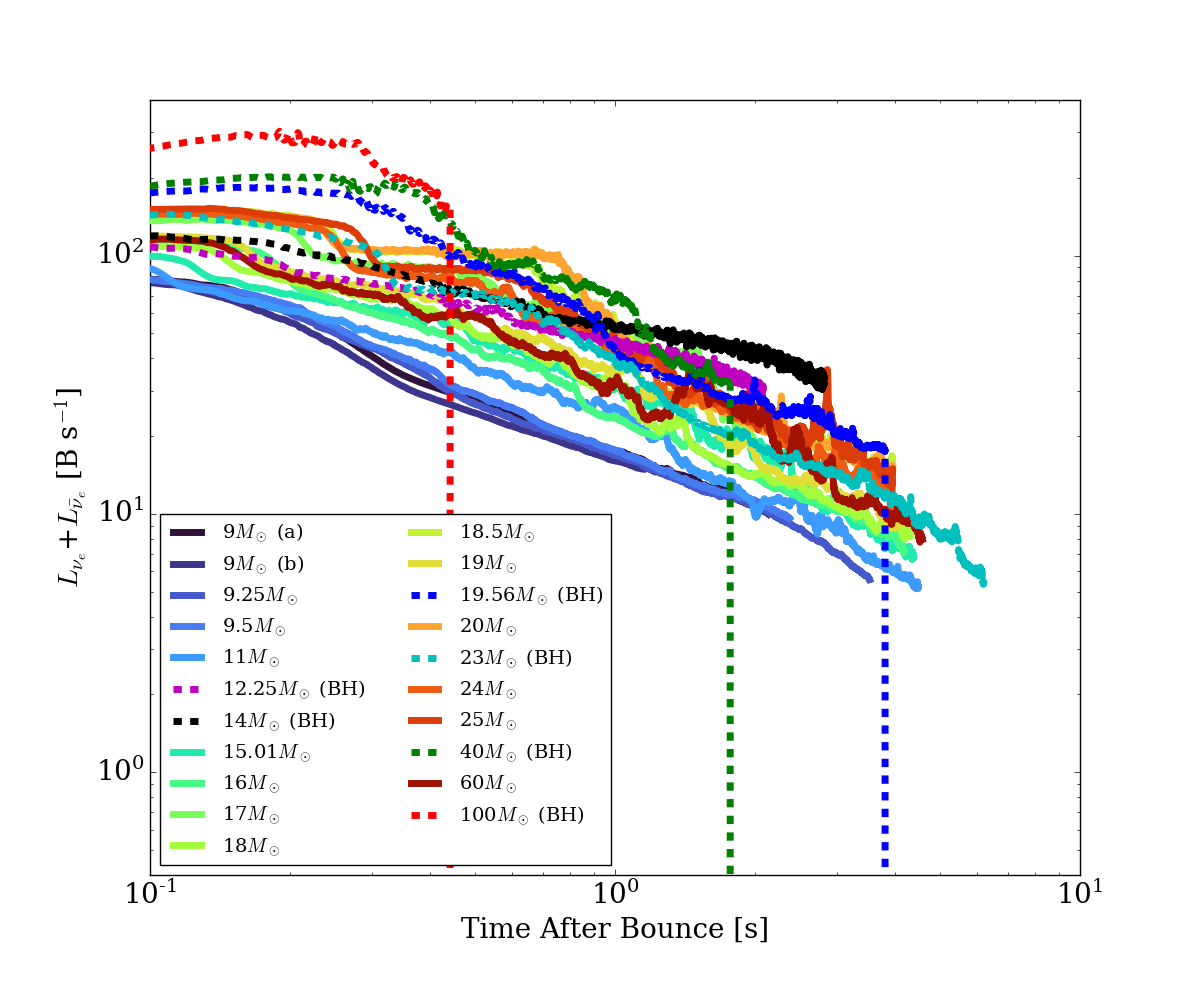}
    \caption{Sum of the solid-angle-integrated $\nu_e$ and $\bar{\nu}_e$ neutrino luminosities (in Bethes per second) for a large collection of our recent 3D CCSN simulations versus the log$_{10}$ of the time after bounce (in seconds). Note that the plots start at 100 milliseconds. The dotted curves are for those models that eventually form black holes and the curves are colored by compactness ($\xi_{1.75}$) from violet to red. The behavior seen here naturally recapitulates that seen in Figure \ref{fig:Qdot} and depicts the near monotonic dependence of luminosity upon compactness. See text for a discussion.}
    \label{fig:lum}
\end{figure}

\begin{figure}
    \centering
    \includegraphics[width=0.47\textwidth]{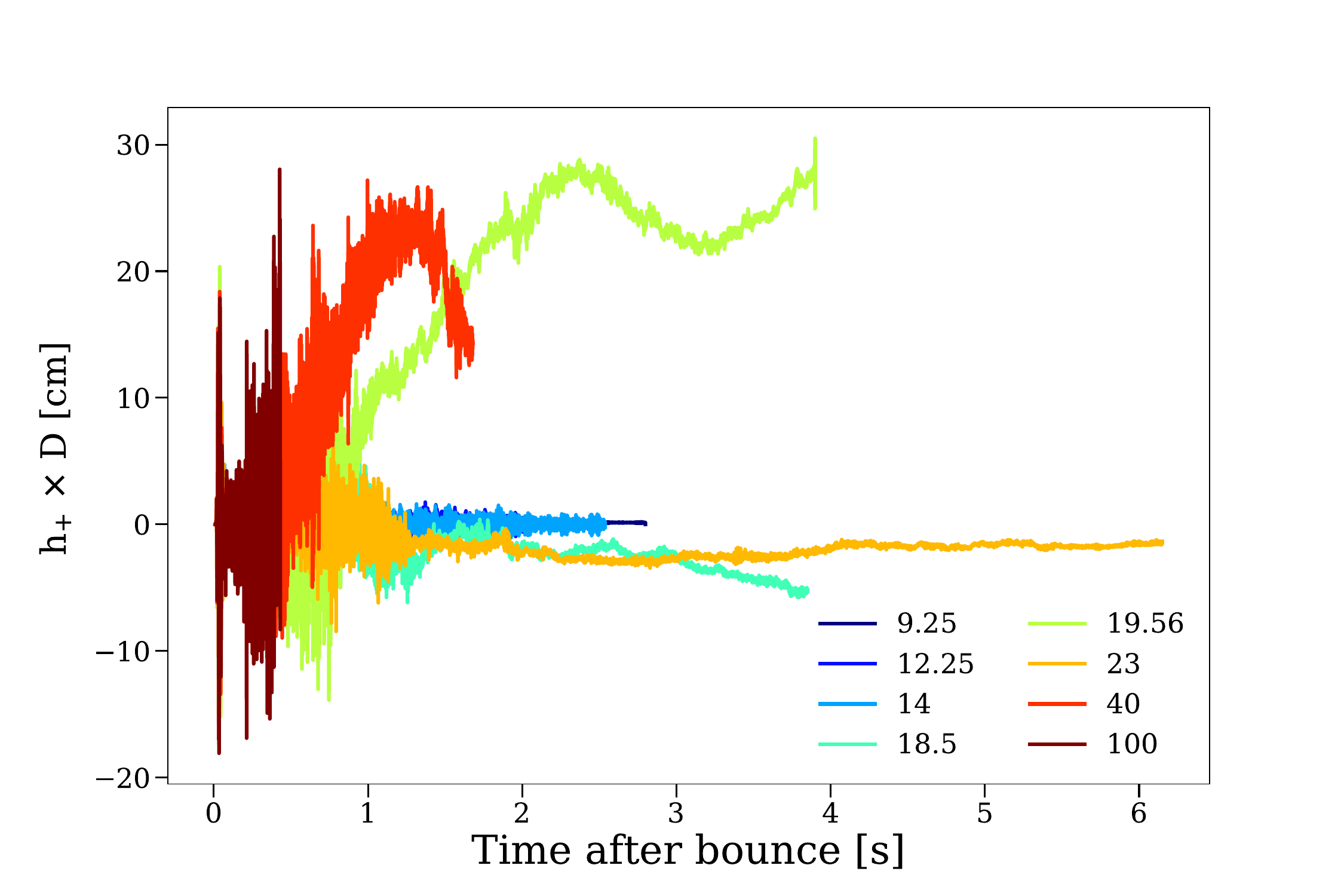}
    \includegraphics[width=0.47\textwidth]{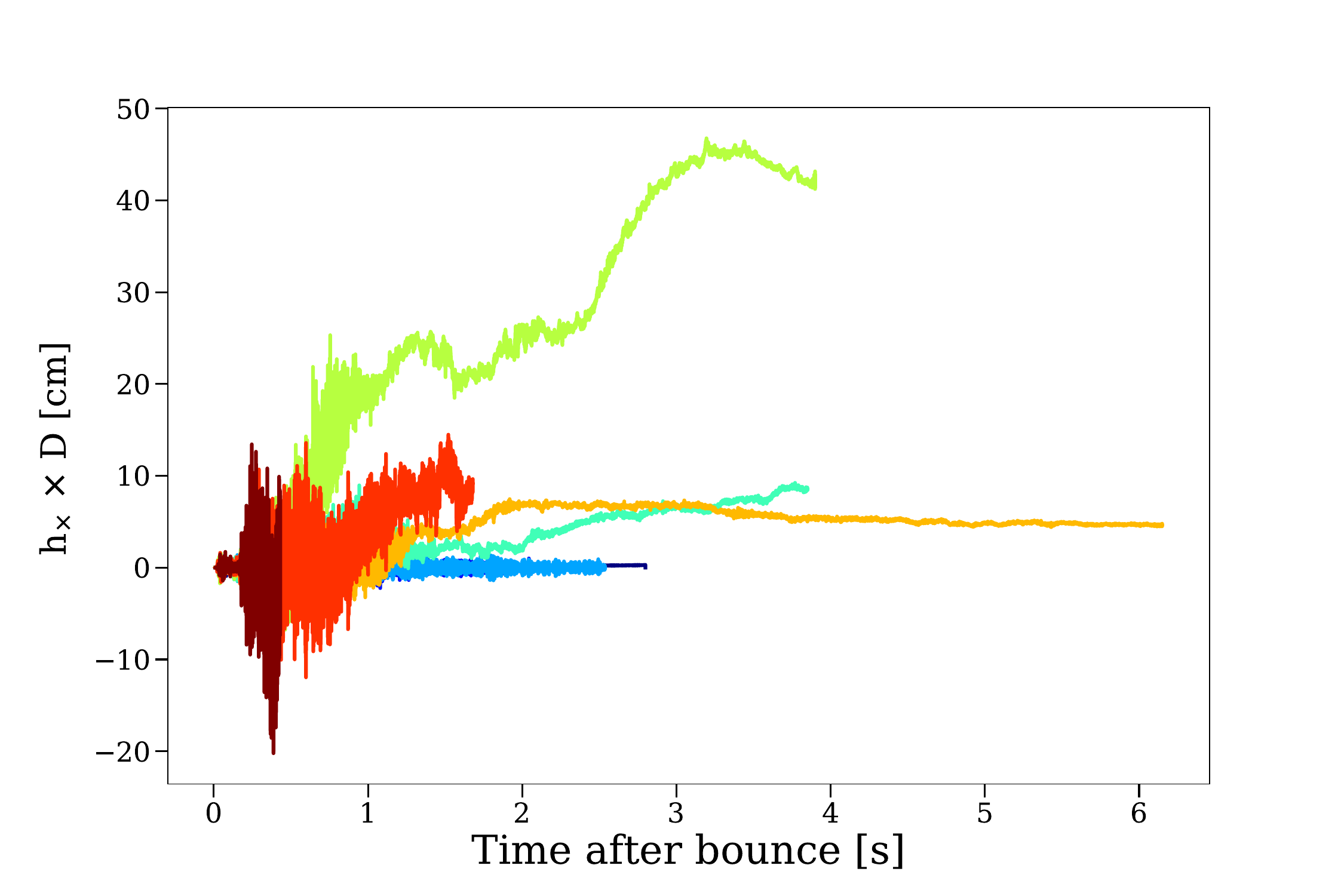}
    \caption{Gravitational wave strain due to matter motions for both $+$ (left) and $\times$ (right) polarizations as a function of time after bounce (in seconds) for our BH models, along with those for two representative non-black-hole formers (9.25-and 18.5-M$_{\odot}$). The increase in strength and the magnitude of the memory offset with initial model compactness and $\dot{M}$ is notable. The greatest strains are achieved by those models that explode and form a black hole. Note the abrupt truncation of the evolution of the strains for those models.  These plots do not include the low-frequency neutrino memory component \citep{choi2024}.}
    \label{fig:gw}
\end{figure}

\subsection{{\bf Channel 4:} Black Hole Formation in the Context of No Supernova Explosion: The ``Silent" Channel}
\label{silent}

Figure \ref{fig:rho_full} identifies two progenitor models from the \citet{swbj16} and \citet{sukhbold2018} collection, the 12.25- and 14-$M_{\odot}$ models, that did not experience shock revival \citep{burrows_40,burrows_correlations_2024}. They have values of $\xi_{1.75}$ in the interval 0.3 to 0.5 where many of the \citet{swbj16} and \citet{sukhbold2018} ZAMS progenitors between $\sim$12 and 15 $M_{\odot}$ reside. In fact, for both 2D and 3D simulations, we have encountered an island in helium-core-mass/compactness space where most of the models fail to explode \citep{burrows_40}. It is only in these intervals that we currently find ``silent" supernovae that would form black holes quiescently\footnote{This might for us be a matter of resolution and we are actively pursuing this idea. It is also possible that hydrodynamic perturbations of significance in this compactness regime could help jump-start these explosions. We are actively pursuing both possibilities, and have already simulated in 3D with perturbations the 12.75-, 13.70-, and 14.43-$M_{\odot}$ models, but to no avail. Nevertheless, this is clearly the most problematic compactness regime and deserves further scrutiny.}. These models have weak silicon/oxygen interface jumps, and this may explain the outcome. As discussed in \S\ref{basics}, they also experience modest post-bounce accretion rates, which if they haven't experienced an early explosion might be a hurdle to subsequent explosion. Simpler analyses \citep{2020ApJ...890..127C} of explodability also conclude that models in this general compactness interval might have difficulty exploding. 

Astronomers do infer a supernova progenitor gap, but above $\sim$18-$M_{\odot}$ \citep{smartt2009b,smith2014,smartt2015,smith2017} and not at these lower masses (however, see \citet{beasor2024_2}) . Importantly, the progenitor mass range in which we see a reticence to explode is contingent in part upon the stellar evolution models we employ. So, due to remaining uncertainties in mass loss rate \citep{smartt2009b}, 3D effects during stellar evolution \citep{fields2020,2021ApJ...908...44Y}, the $^{12}$C($\alpha$,$\gamma$)$^{16}$O rate, shell mergers, overshoot, rotation, and binary interaction \citep{woosley2019,laplace2024}, the mapping between compactness and mass density profile at collapse (of relevance to explodability) and ZAMS mass is very much still in play. This stresses the importance of the initial progenitor model suite one employs. In summary, while there may be a ``silent" channel, it may not reside in the ZAMS mass interval our current 3D CCSN models using the \citet{swbj16} and \citet{sukhbold2018} suggest.

The right panel of Figure \ref{fig:rho_full} shows that 12.25- and 14-$M_{\odot}$ models continue to accrete mass, but at a low rate, while many others around them in compactness (Table \ref{table1}) have asymptoted to a final neutron star mass. We carried these models out in F{\sc{ornax}} to $\sim$2.0$-$3.0 seconds, a longer time than any other model took to explode. It will take minutes to an hour for these models to form black holes. For the 12.25- and 14-$M_{\odot}$ models in Channel 4, the recoil kicks are due solely to net anisotropic neutrino emission and are low, $\sim$6.5 and $\sim$7.0 km s$^{-1}$, respectively. This is calculated under the assumption that all the stellar mass at the time of collapse is eventually accreted and shares in the recoil momentum \citep{burrows_kick_2024}. We suggest, therefore, that Channel 4 black holes will generically experience such low kick speeds, but this conclusion is provisional. 

Figure \ref{fig:rs} shows the behavior of the shock radii for these representives of ``silent" supernovae. In particular, the right hand panel indicates the spiral SASI motion quite clearly as the PNS and shock radii shrink. Figure \ref{fig:lum} indicates that their early neutrino emissions are intermediate, as their compactnesses would suggest. Not shown are their late-time neutrino emissions, which due to continued accretion, should eventually overtake those of the exploding models. Figure \ref{fig:gw} depicts the corresponding gravitational wave signatures, in comparison to those of the other models. We note that for these exemplars of the ``silent" channel there is no matter memory effect, but that the gravitational-wave signal strength is dwarfed by those for the other black hole formation channels. The spiral SASI for the 12.25- and 14-$M_{\odot}$ models continues for the duration of the F{\sc{ornax}} calculations (shown), but should continue much longer (not shown) and will modulate both the gravitational-wave and neutrino signals. Hence, as Figures \ref{fig:lum} and \ref{fig:gw} demonstrate, both their late-time gravitational-wave and neutrino emissions are in principle diagnostics of this modality of black hole formation.  

We speculate that even Channel 4 evolutions would likely not be completely silent. The accretion of mass and only a little angular momentum from either the radius of a giant or a helium core into something only tens of kilometers in radius will result in disks, likely jets, and the associated secondary explosions, however weak. \citet{antoni} find that a turbulent hydrogen envelope should generate upon infall a complex of disks and explode. \citet{nadezhin1980} and \citet{lovegrove2013} have shown that the abrupt alteration in the gravitational potential sensed by the outer envelope matter due to the neutrino burst should generate a shock that would eject mass. In addition, precursor eruptions and ejections of as much as $\sim$0.1 to $\sim$1.0 $M_{\odot}$ have been observed prior to supernovae \citep{kilpatrick2023} and inferred from Type IIp light curves \citep{2015ApJ...814...63M}. Such eruptions should not be related to the details of the CCSN mechanism. Therefore, optical signatures of even this ``silent" channel for black hole formation should be expected upon stellar death. However, such signatures have yet to be observed as such.

\section{Discussion and Conclusions}
\label{conclusions}

What emerges from a comparison of these channels of black hole formation is an interesting and provocative possible systematic trend. We speculate that Channel 2 might give birth to black holes with masses from $\sim$3.0 $M_{\odot}$ to $\sim$10 $M_{\odot}$ in a weak to modest supernova explosion with a small kick that might range from zero to $\sim$100 km s$^{-1}$. Such an event may be difficult to distinguish from a ``regular" supernova. For this channel the energy and black hole mass may be anti-correlated and this might be, but has not been proven to be, a major modality for stellar-mass black hole formation. We emphasize that all the residual black hole and ejecta mass estimates are contingent upon the degree of pre-dynamical mass loss, either in winds or due to binary mass transfer. Both these effects can be significant and will be important factors in the resulting black hole birth mass function. 

In the Channel 1 context, we see high-energy, asymmetrical supernova explosions that leave black holes, but with high kick speeds that suggest they may not stay bound to a companion. The Channel 4 context is ``silent" in that there is no supernova explosion, though there might be a secondary display, and the mass of the final black hole may more closely reflect the mass of the progenitor at collapse. The kicks for this channel due to supernova processes alone are due to net asymmetries in the neutrino emission and are likely to be low. However, they too could range from zero to perhaps $\sim$100 km s$^{-1}$. 

We emphasize that in the high-compactness Channel 1 and Channel 3 black hole formation contexts, as compactness, mantle density, and binding energy increase together they result in countervailing effects. On the one hand, the higher mantle densities and compactnesses result in higher post-bounce mass accretion rates. Higher mass accretion rates result in higher accretion luminosities, which result (along with the associated higher neutrino optical depths in the gain region) in higher neutrino heating rates behind the shock.  Such higher neutrino heating rates ignite explosion. If such high neutrino powers persist long enough, the infant supernova has enough energy to unbind the envelope. However, the high compactnesses and mantle densities that create high heating rates in the first place are also associated with high mass accumulation rates onto the PNS that fatten it quickly. By construction, in this compactness regime the latter is at a rapid rate and a black hole will form. We reemphasize that at high compactness we always witness the rekindling of the stalled shock. The issue is whether the compactness and $\dot{M}$ are so high that the explosion is aborted by the too early formation of a black hole.

Then, the question is whether the time between the inauguration of the supernova and the formation of a black hole at the associated high heating rates is adequate for the integral of the heating rate over time to result in a vigorous enough explosion to unbind most of the outer star. For our solar-metallicity 40-$M_{\odot}$ and 19.56-$M_{\odot}$ models, it was. For our tenth-solar 100-$M_{\odot}$ model, it wasn't.  The results were, on the one hand, \adam{lower mass black holes \citep{shao2022} born in strong, asymmetrical explosions and on the other no CCSN explosion (though in that case a PPISN explosion) and a 37-$M_{\odot}$ black hole}. 

At low metallicity and for ZAMS masses below 100 $M_{\odot}$ (but still high) and possessing high mantle binding energies, a PPISN may not occur, but a large mass black hole is still likely to form. We have yet to simulate such a star, but its ultimate inner behavior may be similar to that seen for our Channel 3 exemplar and its final black hole mass would likely be lower. Clearly, further work is necessary to map this region in stellar phase space.

Hence, Channel 1 black holes would seem to result in an interval, however narrow, of high compactnesses and mantle binding energies. Above this compactness range, a core-collapse supernova is launched, but aborted prematurely, and a much more massive black hole forms (Channel 3). Some of these may be PPISNe. \adam{Lower-metallicity progenitors have higher compactnesses. Hence, we speculate that whether a black hole is formed in one or the other of these channels depends upon metallicity and progenitor mass}, and this highlights the centrality of stellar and binary evolution theory, still very much in flux. We note that the behavior and outcome of our 100-$M_{\odot}$ simulation, in particular vis \`a vis a possible metallicity dependence and the dissipation of the shock wave into a sound wave was anticipated in the earlier papers of \citet{moriya}, \citet{chan2018}, and \citet{rahman2022}.

We reiterate that the dynamics and quantitative outcomes for both Channels 1 and 3 depend upon the nuclear equation of state (EOS). If the maximum mass of a neutron star is higher than we see for the nuclear equation of state we have employed, then the time to collapse to a black hole is longer, with the corresponding alteration in the explosion energies and final black hole masses. This introduces a tantalizing quantitative dependence on the nuclear EOS of these modalities of stellar mass black hole creation that has yet adequately to be explored.

One also asks the question whether there is indeed the fourth black-hole formation channel suggested by the discussion in \S\ref{silent}. This implies a gap, not in black hole mass per se, but in supernova progenitor mass along the supernova progenitor continuum.  We see this near the 12$-$15 $M_{\odot}$ ZAMS-mass region at solar metallicity in an island in helium-core/compactness space \citep{burrows_40} using the \citet{swbj16} and \citet{sukhbold2018} progenitor models. Observers do see a progenitor gap, but not in this interval \citep{smartt2009b,smartt2015,smith2017}. There are numerous caveats to both these theoretical and observational progenitor gaps. On the theory side, the F{\sc{ornax}} and {\it KEPLER} results are clearly not the final word. From the observational side, there are numerous issues yet to resolve involving episodic dust ejection, extinction, and binary effects \citep{beasor2024_2}. In any case, our Channel 4 is the only truly quiescent channel for black hole formation we see among our suite of state-of-the-art 3D CCSN simulations, and even it is unlikely to be truly quiescent. 

Hence, what we find using our current, albeit imperfect, 3D simulation capabilities for the formation channels of stellar mass black holes is not what is commonly envisioned more broadly.  We find four distinct channels for stellar-mass black hole formation in the core-collapse context, two of which clearly lead to supernova explosions. Importantly, the nuclear equation of state, metallicity, and wind and binary mass loss dependencies of these scenarios are important, and require much more scrutiny that we have given them in this paper. In particular, binary interaction can lead to the presence or absence of a hydrogen envelope (is the envelope ``stripped"?). This determines whether there is a reverse shock, which has a direct bearing on the fallback mass and is of direct relevance to the final black hole mass left behind.

Most discussions of stellar-mass black hole formation envision the quiescent formation in a ``failed" supernova of a black hole that will consume all the remaining mass of the progenitor star without display. In this scenario the neutrino and gravitational-wave signals would last seconds and end abruptly \citep{burrows1987,burrows_2000Nature}.
However, what we find currently is that for this more quiescent channel the neutrino signature would last as long as that for neutron star birth \citep{1986ApJ...307..178B} and the gravitational-wave signals would last much longer than that for neutron star birth. 

Whatever survives of our current findings concerning the possible birth contexts of stellar-mass black holes when these important issues are ultimately resolved, the possibility of more exotic channels of black hole formation, and the possibility of accompanying supernovae in a subset, suggest that this subject may be much richer than previously envisioned.

\section*{Data Availability}  

The data presented in this paper can be made available upon reasonable request to the first author.  

\section*{Acknowledgments}

We thank Stan Woosley for providing his 100-$M_{\odot}$ model prior to publication and Christopher White, Matthew Coleman, and Benny Tsang for long-term productive interactions. DV acknowledges support from the NASA Hubble Fellowship Program grant HST-HF2-51520. AB acknowledges former support from the U.~S.\ Department of Energy Office of Science and the Office of Advanced Scientific Computing Research via the Scientific Discovery through Advanced Computing (SciDAC4) program and Grant DE-SC0018297 (subaward 00009650) and former support from the U.~S.\ National Science Foundation (NSF) under Grant AST-1714267. We are happy to acknowledge access to the Frontera cluster (under awards AST20020 and AST21003). This research is part of the Frontera computing project at the Texas Advanced Computing Center \citep{Stanzione2020}. Frontera is made possible by NSF award OAC-1818253. Additionally, a generous award of computer time was provided by the INCITE program, enabling this research to use resources of the Argonne Leadership Computing Facility, a DOE Office of Science User Facility supported under Contract DE-AC02-06CH11357. Finally, the authors acknowledge computational resources provided by the high-performance computer center at Princeton University, which is jointly supported by the Princeton Institute for Computational Science and Engineering (PICSciE) and the Princeton University Office of Information Technology, and our continuing allocation at the National Energy Research Scientific Computing Center (NERSC), which is supported by the Office of Science of the U.~S.\ Department of Energy under contract DE-AC03-76SF00098.

\newpage

\clearpage

\bibliographystyle{aasjournal}
\bibliography{References}

\label{lastpage}
\end{document}